\documentclass[a4paper,twocolumn,showpacs]{revtex4}
\usepackage{dcolumn}
\usepackage{amsmath}
\usepackage{graphicx}

\providecommand{\gtrsim}{\:\raisebox{.25ex}{$>$}\hspace*{-.75em}
\raisebox{-.93ex}{$\sim$}\:}
\providecommand{\lesssim}{\:\raisebox{.25ex}{$<$}\hspace*{-.75em}
\raisebox{-.93ex}{$\sim$}\:}

\begin{document}
\title{Test ion acceleration in the field of expanding planar electron
cloud }
\author{M.M.~Basko}
\email{basko@itep.ru}

\affiliation{Institute for Theoretical and Experimental Physics,
Moscow, Russian Federation}


\begin{abstract}
New exact results are obtained for relativistic acceleration of test
positive ions in the laminar zone of a planar electron sheath evolving
from an initially mono-energetic electron distribution. The electron
dynamics is analyzed against the background of motionless foil ions.
The limiting gamma-factor $\gamma_{p\infty}$ of accelerated ions is
shown to be determined primarily by the values of the ion-electron
charge-over-mass ratio $\mu =m_eZ_p/m_p$ and the initial gamma-factor
$\gamma_0$ of the accelerated electrons. For $\mu> \frac{1}{8}$ a test
ion always overtakes the electron front and attains $\gamma_{p\infty}>
\gamma_0$. For $\mu< \frac{1}{8}$ a test ion can catch up with the
electron front only when $\gamma_0$ is above a certain critical value
$\gamma_{cr}$, which for $\mu \ll 1$ can most often be evaluated as
$\gamma_{cr} = \frac{1}{4} \mu\exp\left(\mu^{-1}-1\right)$. In reality
the protons and heavier test ions, for which $\gamma_{cr}> 10^{398}$ is
enormous, always lag behind the front edge of the electron sheath and
have $\gamma_{p\infty}< \gamma_0$; for their maximum energy an
appropriate intermediate asymptotic formula is derived. The domain of
applicability of the laminar-zone results is analyzed in detail.

\end{abstract}

\pacs{52.30.Ex, 52.38.Kd, 52.40.Kh}

\maketitle

\section{Introduction}

One of the most impressive latest achievements in laser-plasma
interaction has been the observation of well-collimated high-energy
proton and ion beams produced from thin metallic foils irradiated by
ultra-intense sub-picosecond laser pulses
\cite{ClKr.00,MaGu.00,SnKey.00,HeKa.02}. An exceptional quality,
demonstrated recently for thus produced proton beams \cite{CoFu.04},
makes them very promising for many potential applications
\cite{MoTa.06}.

In its gross features, the mechanism of ion acceleration in the cited
and other similar experiments is believed to be reasonably well
understood, and has been nicknamed TNSA (target normal sheath
acceleration) \cite{HaBr.00,WiLa.01}. High directionality and low phase
volume of generated protons \cite{CoFu.04} indicate that they are
accelerated in a highly ordered electric field normal to the virtually
unperturbed planar rear surface of the laser-irradiated foil. The
electric field is caused by charge separation in the sheath layer that
is formed by energetic electrons generated by absorption of the laser
pulse.

However, an attempt to provide a more detailed and comprehensive
theoretical description leads to a very complex system of plasma
dynamics equations that can hardly be ever solved rigorously. To
establish practically useful dependences and relations, one has to
introduce additional simplifications. As is typical in other areas of
physics, of particular value for gaining a deeper understanding of the
process of ion acceleration prove to be certain particular idealized
but exactly solvable problems. Salient and well known examples are
(i)~a~self-similar evolution of the ion distribution function by
quasi-neutral plasma expansion into vacuum \cite{GuPa.65}, and
(ii)~a~virtually exact two-fluid solution of the isothermal plasma
expansion with a full account of charge separation effects
\cite{Mora03}.  In this paper we present a rigorous solution and a full
parametric analysis of another such idealized problem, namely, the
problem of test ion acceleration in a dynamic sheath of relativistic
electrons with the delta-function initial velocity distribution.

Typically, fast protons in laser irradiated metallic foils originate
from a thin (few nanometers) contaminant layer of water and
hydrocarbons at the foil surface \cite{GiJo.86,SnKey.00,HeKa.02}. Then,
a natural simplification would be to assume that the heavy bulk ions
(like Au for example) of a metallic foil are infinitely heavy and stay
at rest, while the accelerated protons are treated as test positive
charges initially located at the foil surface. Our present solution is
essentially based on this assumption.

Next, one has to choose how to treat the electrons. A widely used
assumption is that the electrons instantaneously relax to the
equilibrium Boltzmann distribution in the time-dependent electrostatic
potential of the expanding plasma: it was employed in both solutions
\cite{GuPa.65,Mora03} cited above. With immobile ions, such an
assumption allows straightforward calculation of the electrostatic
sheath potential $\phi(x)$ either in the one-temperature \cite{CrAu.75}
or multi-temperature \cite{PaTi.04} cases. An obvious problem with this
approximation is that it leads to a diverging result for the maximum
energy of accelerated test ions because the corresponding potential
$\phi(x) = -(2T/e) \ln\left[1+x/(\sqrt{2}\lambda_D)\right]$
\cite{CrAu.75} logarithmically diverges at $x\to +\infty$ in the planar
geometry; here $\lambda_D$ is the Debye length at the base of the
electron sheath with temperature $T$, $+e$ is the elementary charge.
However, as was pointed out by Gurevich \textit{et al.} \cite{GuPa.65},
this divergence is not physical because even if one assumes that the
hot electrons have a perfect Maxwellian distribution initially, at
$t=0$, it still takes an increasingly long time $\Delta t \simeq
t_{eq}(x) \propto x$ for the Boltzmann relation $n_e \propto
\exp(e\phi/T)$ to establish at an increasingly large distance $x$ from
the initial plasma surface (for more details see section \ref{s:rlxL0}
below). As possible remedies, attempts have been made to use ad hoc
quasi-equilibrium electron distributions truncated either at high
velocities \cite{PeMo78,KiMi.83} or at large distances \cite{PaLo04}.
Evidently, neither of these two approaches is fully self-consistent.

Without the Boltzmann relation, a self-consistent treatment requires
that one starts with a given initial electron distribution function at
$t=0$, and then calculates its evolution for $t>0$. For high-energy
(multi-MeV) electrons this can be done in the collisionless
approximation. In this work we solve this problem in the simplest case
of initially monoenergetic electrons, i.e.\ when at $t=0$ all the free
electrons of a uniform planar foil have one and the same initial
velocity $v_0$ perpendicular to the foil. Rigorous results are obtained
for test ion acceleration in the outer laminar zone (for strict
definition see section~\ref{s:eno} below) of the dynamically evolving
electron sheath. Particular attention is paid to the limiting energy of
accelerated ions at $t \to \infty$, which is always finite within the
adopted model.

This paper is not the first publication addressing thus formulated
problem: to a significant extent, it builds upon earlier work by
Bulanov \textit{et al.} \cite{BuEs.04}. The new progress made here
includes the following key issues. Temporal behavior of the boundary
between the outer laminar and the inner relaxation zones of the
dynamically evolving electron sheath is studied in detail.
Consequently, the domain of applicability of the laminar-zone results
for test ions is clearly identified in the full parameter space of the
problem. In contrast to Ref.~\cite{BuEs.04}, the electric field in the
laminar zone is calculated exactly and not to the accuracy of the
linear in $x$ term. It is proven that the answer to the intriguing
question of whether a test ion can overtake the electron front (and,
consequently, surpass the initial electron velocity $v_0$) is
determined by a critical relationship between the initial gamma-factor
$\gamma_0$ of accelerated electrons and the ion-electron
charge-over-mass ratio $\mu$ [defined in Eq.~(\ref{3par=}) below]. A
fully relativistic intermediate asymptotic formula
(\ref{iL0:Ias-Pi_p=}) is derived which may be used in realistic
situations to evaluate the maximum energy of accelerated protons and
heavier ions.

\section{Formulation of the problem}

\begin{figure}[htb]
\includegraphics[width=50mm]{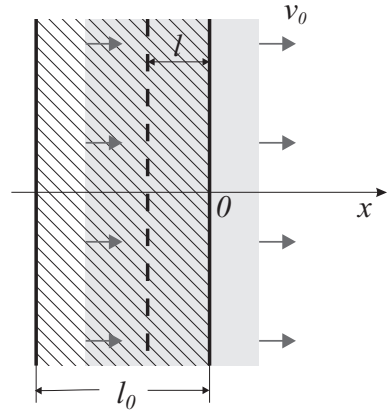}
\caption{Schematic view of a plasma foil with motionless bulk ions
(hatched area) and electrons (grey area) boosted to a velocity $v_0$.
Dashed vertical line marks the initial position of electrons with the
Lagrangian coordinate $l$.} \label{f:1}
\end{figure}

Consider a uniform electrically neutral plasma foil of thickness $l_0$
with an initial density of free electrons $n_0$.  At time $t=0$ all the
free electrons are set in motion with the same initial velocity $v_0$
perpendicular to the foil (see Fig.~\ref{f:1}); $v_0$ can be
arbitrarily close to the speed of light $c$. At later times $t>0$ the
motion of electrons, treated as a collisionless charged fluid, is
governed by the electric field $E(t,x)$, which arises due to charge
separation in the evolving plasma cloud. The origin of the $x$-axis,
directed along the initial electron velocity $\vec{v}_0$, is chosen at
the forward foil surface, so that initially the foil occupies the
region $-l_0 \leq x \leq 0$. The bulk foil ions are assumed to be
infinitely heavy and staying at rest. Our goal is to calculate the
motion of a test ion of charge $+eZ_p$ and mass $m_p$ placed  initially
at the foil surface $x=0$, which is accelerated by the electric field
$E(t,x)$ in the positive direction of the $x$ axis. Note that here and
below $m_p$ is not necessarily the proton mass.

It is easy to understand that this problem is governed by only three
independent dimensionless parameters, which we choose to be
\begin{equation}\label{3par=}
  \mu =\frac{m_e Z_p}{m_p}, \quad \Lambda = \frac{l_0\omega_0}{v_0
  \sqrt{\gamma_0}}, \quad \gamma_0 =(1-\beta_0^2)^{-1/2};
\end{equation}
here $m_e$ is the electron mass,
\begin{equation}\label{omega_0=}
  \omega_0 =\left(\frac{4\pi e^2 n_0}{m_e} \right)^{1/2}
\end{equation}
is the plasma frequency in the initial configuration, $e$ is the
positive elementary charge, and $\beta_0 =v_0/c$. If a proton (or some
other light ion) is chosen as a test particle, the charge-over-mass
ratio $\mu$ is small, $\mu \leq 1/1836 \ll 1$. We will, however,
explore the entire possible range of $0 < \mu < \infty$, firstly, for
the sake of completeness of the analysis, and, secondly, keeping in
mind that light positive particles --- such as positrons, $\pi^+$ or
$\mu^+$ mesons --- may in principle be created and accelerated under an
intense laser irradiation.

For a non-thermal electron cloud considered here the quantity $v_0
\sqrt{\gamma_0} /\omega_0$ plays a role of the Debye length. Then, the
parameter $\Lambda$ is the initial foil thickness in units of the Debye
length. Of particular interest is the limit $\Lambda =0$ of a
geometrically infinitely thin foil which, however, has a finite number
\begin{equation}\label{Sigma=}
  \Sigma =n_0l_0
\end{equation}
of electrons per unit surface area. On the one hand, the effects of
charge separation are the strongest in this limit for a given value of
$v_0$. On the other hand, it is in this limit that most of the exact
results can be established analytically; many of them are then
straightforwardly extended to a more general case of $\Lambda \gtrsim
1$.

Parameter $\gamma_0$ is the relativistic gamma-factor of the
accelerated electrons. In the non-relativistic limit of $\beta_0 \ll
1$, when $\gamma_0 \approx 1 +\frac{1}{2} \beta_0^2$, this parameter
becomes irrelevant, and we are left with only two principal parameters
$\mu$ and $\Lambda$. On a par with $\beta_0$ and $\gamma_0$, the
dimensionless momentum $\Pi_0 = \beta_0 \gamma_0$ is used below as the
principal kinematic characteristic of the accelerated electrons.

\section{Motion of electrons}\label{s:em}

\subsection{General notation and relationships}\label{s:eno}

Motion of electrons is described by a function $x_e(t,l)$ [or by a
function $x_e(t,\xi)$], where $l$ (or $\xi$) is a Lagrangian coordinate
in the electron fluid: $x_e(t,l)$ is the position at time $t$ of an
electron whose original position at $t=0$ was $x_e(0,l) = -l$ (the
broken line in Fig.~\ref{f:1}). The dimensionless Lagrangian coordinate
$0\leq \xi \leq 1$ is defined as
\begin{equation}\label{em:xi=}
  \xi= \frac{l}{l_0}.
\end{equation}
Then, the front edge of the electron cloud is at $\xi= 0$, its rear
edge is at $\xi =1$.

At any fixed time $t$ one can invert $x_e(t,\xi)$ with respect to $\xi$
to obtain the inverse function $\xi = \xi(t,x)$ defined inside the
expanding electron cloud. Without electron-electron collisions the
function $\xi(t,x)$ ceases to be single-valued with respect to $x$
after some time --- even if it were so initially. In this paper we use
the term ``laminar zone'' for that region of the electron cloud where
$\xi(t,x)$ is single-valued (see Figs.~\ref{f:2} and \ref{f:3} below).
In the remaining ``relaxation zone'', where $\xi(t,x)$ is multi-valued,
electrons gradually relax to the equilibrium Boltzmann distribution.

The function $x_e(t,\xi)$ is found by solving the following equations
of electron motion
\begin{subequations} \label{em:dx,p_e/dt=}
\begin{eqnarray}\label{em:dx_e/dt=}
  \frac{1}{c}\frac{dx_e}{dt} &=&
  \frac{\Pi_e}{\sqrt{1+\Pi_e^2}} \equiv\beta_e, \\ \label{em:dp_e/dt=}
  \frac{d\Pi_e}{dt} &=& -\frac{e}{m_ec} \, E(t,x_e),
\end{eqnarray}
\end{subequations}
where  $\beta_e = v_e/c$, and $\Pi_e = \beta_e \gamma_e = \beta_e
(1-\beta_e^2)^{-1/2}$; the Lagrangian time derivative $d/dt$ is
calculated at a fixed $\xi$. As is well known, equations
(\ref{em:dx,p_e/dt=}) admit the energy integral
\begin{equation}\label{em:en-int=}
  m_ec^2\Sigma \int\limits_0^1 \gamma_e(t,\xi)\, d\xi +\frac{1}{8\pi}
  \int\limits_{-\infty}^{+\infty} E^2\, dx= m_ec^2\Sigma\, \gamma_0.
\end{equation}

The electric field $E(t,x)$ is obtained by solving the Poisson
equation, which in the planar geometry of our problem yields
\begin{equation}\label{em:E=}
  E(t,x) = 4\pi e \Sigma \left[\sigma_e(t,x) -\sigma_i(x)\right];
\end{equation}
here $0 \leq \sigma_e(t,x) \leq 1$ [$0 \leq \sigma_i(x) \leq 1$] is the
fraction of the total number of electrons (ions) above $x$ at time $t$,
i.e.\
\begin{equation}\label{em:sig_e_def=}
  \sigma_{e}(t,x) = \frac{1}{\Sigma} \int\limits_x^{\infty}
  n_{e}(t,x')\, dx'.
\end{equation}
For motionless ions one obviously has
\begin{equation}\label{em:sig_i=}
  \sigma_i(x) = \left\{ \begin{array}{ll}
  1, & x< -l_0, \\ -x/l_0, & -l_0 \leq x\leq 0, \\
  0, & x> 0.  \end{array}\right.
\end{equation}
Calculation of $\sigma_e(t,x)$ depends on whether $x$ happens to be in
the laminar or relaxation zone of the electron cloud. In the laminar
zone one simply has $\sigma_e(t,x) =\xi(t,x)$, and equations of motion
(\ref{em:dx,p_e/dt=}) can be solved analytically.

In the relaxation zone, where the function $\xi(t,x)$ becomes
multi-valued with respect to $x$, one has to use a more general
expression
\begin{equation}\label{em:sig_e=}
  \sigma_e(t,x) = \sum_{x_e(t,\xi) >x} \Delta\xi,
\end{equation}
where summation is done over all the segments $\Delta\xi$ which at time
$t$ are located inside the interval $(x,+\infty)$. Clearly, if
Eqs.~(\ref{em:dx,p_e/dt=}) are to be solved with a full account of the
relaxation zone, it can only be done numerically. To this end, a
numerical code TIAC (\underline{t}est \underline{i}on
\underline{ac}celeration) has been written, which calculates the
electron motion and the ensuing electric field $E(t,x)$ with a full
account of possible mutual interpenetration of different elements of
the electron fluid. Because of rapid randomization (for details see
subsection~\ref{s:rlxL0}) of the electron motion at the core of the
relaxation zone, such straightforward calculations can only be realized
within a limited time span of the order of 100 periods of electron
oscillations near the foil surface.

\subsection{Solution for $\Lambda =0$} \label{s:eL0}

Here we consider the limit $l_0 \to 0$ of a geometrically very thin
foil with a finite value of electron number per unit area $\Sigma
=n_0l_0$, where both the initial electron density $n_0$ and the plasma
frequency $\omega_0$ become formally infinite. In this case it is
convenient to introduce the following units of time and length
\begin{equation}\label{eL0:[t]=}
  [t]=t_{\Sigma} \equiv \frac{m_ev_0}{4\pi e^2\Sigma}\, \gamma_0, \quad
  [x]=l_{\Sigma} \equiv v_0t_{\Sigma},
\end{equation}
which replace the usual time and length scales
$\sqrt{\gamma_0}/\omega_0$, $v_0\sqrt{\gamma_0}/\omega_0$ in a
finite-density plasma. Evidently, the length unit $l_{\Sigma}$ plays a
role of the Debye length in our dynamic electron sheath. Below, the
quantities measured in units (\ref{eL0:[t]=}) are marked with a bar.
The electron trajectories are represented by a two-parameter family of
curves $\bar{x}_e(\bar{t},\xi,\beta_0)$.

\subsubsection{Electron trajectories}

\begin{figure}[htb]
\includegraphics[width=75mm]{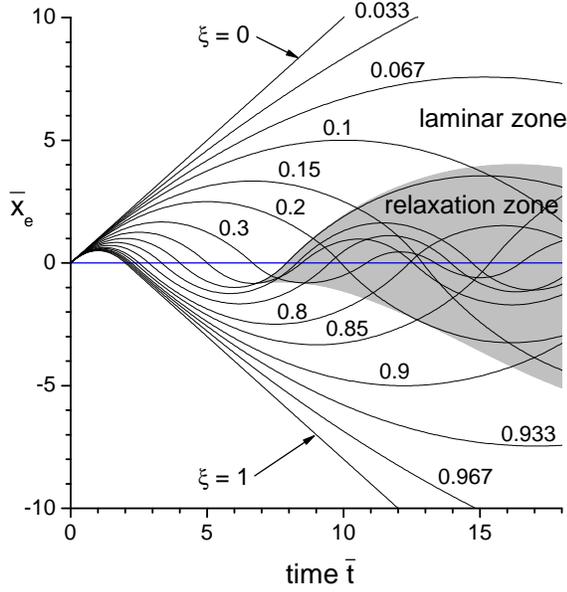}
\caption{Electron trajectories for $\Lambda=0$ in the non-relativistic
limit of $\Pi_0 \ll 1$. Each curve is marked by the corresponding $\xi$
value. The relaxation zone is shown as a grey shaded area.} \label{f:2}
\end{figure}

In the laminar zone, where $\sigma_e =\xi$, Eqs.~(\ref{em:dx,p_e/dt=})
are solved analytically. The two relevant branches of this solution,
obtained with the initial conditions $\bar{x}_e(0,\xi,\beta_0) =0$,
$\Pi_e(0,\xi,\beta_0) =\Pi_0 \equiv \beta_0\gamma_0$, are
\begin{subequations}\label{eL0:Pi,bx_e-1=}
\begin{eqnarray}\label{eL0:Pi_e-1=}
  \Pi_e &=& \Pi_0 (1-\xi\bar{t}), \\ \label{eL0:bx_e-1=}
  \bar{x}_e &=& \frac{\bar{t}(2-
  \xi\bar{t})}{1+\sqrt{1-\beta_0^2\xi \bar{t}(2-\xi\bar{t})}},
\end{eqnarray}
\end{subequations}
for $0 <\bar{t} < 2/\xi$, and
\begin{subequations}\label{eL0:Pi,bx_e-2=}
\begin{eqnarray}\label{eL0:Pi_e-2=}
  \Pi_e &=& \Pi_0 \left[(1-\xi)\bar{t} +1 -2/\xi\right],
  \\ \label{eL0:bx_e-2=}
  \bar{x}_e &=& \frac{\bar{t}^2(1-\xi) +2\bar{t}(1-
  2/\xi) +4/\xi^2}{1+\sqrt{1-\beta_0^2\xi \bar{t}(2-\xi\bar{t})}},
\end{eqnarray}
\end{subequations}
for $2/\xi <\bar{t} < 2/\left[\xi(1-\xi)\right]$.

\noindent In the upper half-space $\bar{x}_e> 0$
Eq.~(\ref{eL0:bx_e-1=}) is easily inverted with respect to $\xi$, which
leads us to the following expression for the electric field
\begin{equation}\label{eL0:E=}
  E(\bar{t},\bar{x}) =4\pi e\Sigma\, \xi(\bar{t},\bar{x})= 8\pi e\Sigma\,
  \frac{\bar{t}- \bar{x}}{\bar{t}^2 -
  \beta_0^2\bar{x}^2}.
\end{equation}
Inside the relaxation zone, where the electron trajectories intersect
with one another, one has to abandon
Eqs.~(\ref{eL0:Pi,bx_e-1=})--(\ref{eL0:E=}) and solve
Eqs.~(\ref{em:dx,p_e/dt=}) numerically to calculate
$\bar{x}_e(\bar{t},\xi,\beta_0)$ and $E(\bar{t},\bar{x})$.

\begin{figure}[htb]
\includegraphics[width=75mm]{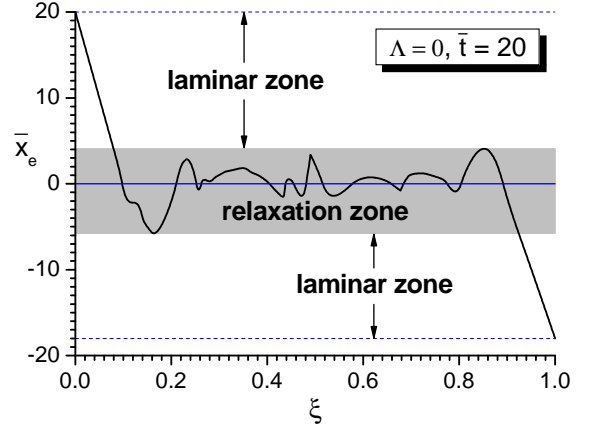}
\caption{Position of individual electrons $\bar{x}_e(\bar{t},\xi)$ in
the non-relativistic case at time $\bar{t} =20$. Without
electron-electron collisions,  $\bar{x}_e(\bar{t},\xi)$ is
non-monotonic with respect to the Lagrangian coordinate $\xi$ in the
relaxation zone (shaded area). The number of local minima and maxima
along $\xi$ rapidly (exponentially) increases with time.} \label{f:3}
\end{figure}

In the non-relativistic limit $\bar{x}_e(\bar{t},\xi)
=\bar{x}_e(\bar{t},\xi,0)$ is a universal function of two variables
$\bar{t}$ and $\xi$. It is plotted in Fig.~\ref{f:2} for a selection of
$\xi$ values as calculated with the TIAC code. In Fig.~\ref{f:3} the
non-relativistic function $\bar{x}_e(\bar{t},\xi)$ is plotted versus
$\xi$ for $\bar{t} =20$. Qualitatively, the relativistic trajectories
$\bar{x}_e(\bar{t},\xi,\beta_0)$ look similar to
$\bar{x}_e(\bar{t},\xi,0)$.  In particular, for any $0< \beta_0 <1$ the
rear edge of the electron cloud $\xi=1$ turns around (i.e.\ has
$\Pi_e=0$) at $\bar{t} =1$, and later crosses the foil at $\bar{t} =2$
with $\Pi_e = -1$. During the period $0< \bar{t}< 2$ there exists a
vacuum gap between the foil ions and the ejected electrons. After its
closure, the front and the rear  edges of the electron cloud propagate
freely in opposite directions with constant velocities $\pm v_0$.

Numerical solution of Eqs.~(\ref{em:dx,p_e/dt=}) enables one to trace
the onset and subsequent expansion of the relaxation zone, shown in
Figs.~\ref{f:2} and \ref{f:3} as shaded areas. In this zone
oscillations of electrons around the positively  charged foil,
occurring on a time scale $t_{\Sigma}$, rapidly become  stochastic and
ultimately lead to establishment of  the Maxwell-Boltzmann distribution
with a certain temperature that can be found from the energy integral
(\ref{em:en-int=}). The starting point of the relaxation zone can be
calculated analytically by applying the conditions
$\partial\bar{x}_e/\partial\xi =0$, $\partial^2\bar{x}_e/\partial\xi^2
=0$ to Eq.~(\ref{eL0:bx_e-2=}); in the non-relativistic limit it has
the coordinates
\begin{equation}\label{eL0:nr-bt_r0=}
  \bar{t}_{r0} = \frac{27}{4}, \quad \xi_{r0} =\frac{4}{9},
  \quad \bar{x}_{r0} = -\frac{27}{32},
\end{equation}
which become
\begin{equation}\label{eL0:ur-bt_r0=}
  \bar{t}_{r0} = 6, \quad \xi_{r0} =\frac{1}{2},
  \quad \bar{x}_{r0} = -2.
\end{equation}
in the ultra-relativistic limit $\gamma_0 \gg 1$.

\subsubsection{Evolution of the relaxation zone} \label{s:rlxL0}

\begin{figure}[htb]
\includegraphics[width=75mm]{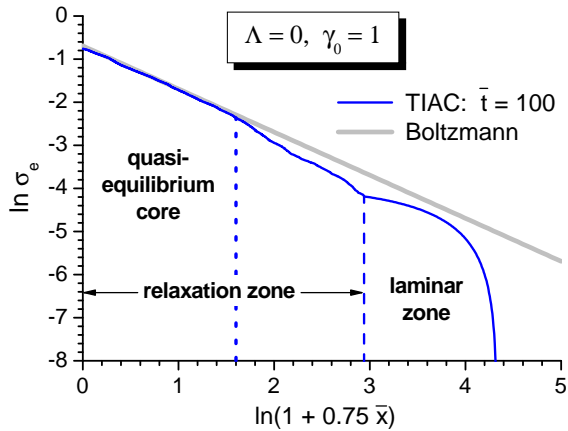}
\caption{Spatial profile of the electron column density
$\sigma_e(\bar{t},\bar{x})$ as calculated numerically with the TIAC
code in the non-relativistic limit for $\bar{t}=100$. The equilibrium
profile (\ref{eL0:sig_eq=}) obtained by using the Boltzmann relation is
shown as a thick grey straight line.} \label{f:4}
\end{figure}

Even though electrons are treated as a collisionless fluid, the
presence of a positively charged ion sheet is sufficient for the
electron subsystem to develop a stochastical behavior. Once the
electron trajectories in the central region begin to intersect one
another in the process of oscillations across the ion layer (see
Fig.~\ref{f:2}), their further motion becomes increasingly stochastic.
As a result, an isothermal quasi-equilibrium core gradually develops
inside the relaxation zone, where the electron density obeys the
Boltzmann relation $n_e(x) \propto \exp\left(e\phi/T\right)$; here
$\phi(x)$ is the equilibrium electrostatic potential, and $T$ is the
temperature of the quasi-equilibrium core. This qualitative picture is
fully confirmed by the results of direct simulations with the TIAC code
presented in Fig.~\ref{f:4}.

Having adopted the Boltzmann relation, one easily solves the Poisson
equation and obtains
\begin{equation}\label{eL0:bar_phi=}
  \bar{\phi}(\bar{x}) = -2\bar{T}\, \ln\left(1+
  \frac{|\bar{x}|}{4\bar{T}}\right),
\end{equation}
where $\bar{\phi}$ and $\bar{T}$ are, respectively, the potential in
units $m_ev_0^2 \gamma_0/e$ and the temperature in units $m_ev_0^2
\gamma_0$. The temperature $\bar{T}$ is found from the energy integral
(\ref{em:en-int=}), which in our case transforms to a transcendental
equation
\begin{equation}\label{eL0:T=}
  2\bar{T} \beta_0^2 +\frac{
  K_0\left(1/\beta_0\Pi_0\bar{T}\right)}{\gamma_0\,
  K_1\left(1/\beta_0\Pi_0\bar{T}\right)} =1;
\end{equation}
here $K_0(z)$ and $K_1(z)$ are the modified Bessel functions. The
limiting values of $\bar{T}$ are
\renewcommand{\arraystretch}{2.0}%
\begin{equation}\label{eL0:T_lim=}
  \bar{T} = \left\{ \begin{array}{ll} \displaystyle \frac{1}{3}, &
  \Pi_0 \ll 1, \\ \displaystyle
  \frac{1}{2}, & \Pi_0 \gg 1. \end{array} \right.
\end{equation} \renewcommand{\arraystretch}{1.0}%
Figure~\ref{f:4} compares the equilibrium electron column density [as
defined by Eq.~(\ref{em:sig_e_def=})]
\begin{equation}\label{eL0:sig_eq=}
 \left. \sigma_{e,eq}(\bar{x})\right|_{\bar{x}>0} =
  \frac{1}{2} \left(1+ \frac{\bar{x}}{4\bar{T}}\right)^{-1},
\end{equation}
calculated from Eq.~(\ref{eL0:bar_phi=}), with that obtained from the
TIAC numerical simulations in the non-relativistic limit for $\bar{t}
=100$. One clearly distinguishes a quasi-equilibrium core of the
relaxation zone, which is adequately described by the Boltzmann
relation. Departures from the equilibrium are significant in the outer
part of the relaxation zone and, of course, in the laminar zone.

To assess practical applicability of our results for test ion
acceleration in the laminar zone, obtained below, one needs to know how
the boundary $x_r(t)$ between the relaxation and the laminar zones
evolves in time. We analyze this evolution by combining an analytical
estimate in the limit of $t \to \infty$ with the TIAC simulations for
times $\bar{t} \lesssim 100$.

Since randomization of the electron motion in the relaxation zone leads
to establishment of the Maxwell-Boltzmann distribution, we can invoke
the following argument to evaluate the width $\bar{x}_r$ of this zone
at times $\bar{t} \gg 1$: by a time  $\bar{t}$ the relaxation zone
spreads to a distance $\bar{x}_r$ such that $\bar{t} = N_r\,
\bar{t}_{eq}(\bar{x}_r)$, where $N_r$ is a numerical factor of the
order unity, and $\bar{t}_{eq}(\bar{x}_r)$ is the travel time between
$\bar{x}=0$ and $\bar{x}= \bar{x}_r$ in the equilibrium potential
(\ref{eL0:bar_phi=}) of an electron whose kinetic energy vanishes at
$\bar{x}= \bar{x}_r$. In a sense, $t_{eq}(x_r)$ is a timescale on which
the information about maxwellization of the velocity distribution up to
a certain limiting value $v_r$ is transferred from $x=0$ to the
corresponding limiting distance $x_r$ in the infinite potential well
(\ref{eL0:bar_phi=}). This argument sounds perfectly reasonable for the
quasi-equilibrium core, and may be surmised to apply to the entire
relaxation zone as well.

The time $\bar{t}_{eq}(\bar{x}_r)$ is given by an integral
\begin{equation}\label{eL0:t_eq_def=}
 \bar{t}_{eq}(\bar{x}_r) = \int\limits_0^{\bar{x}_r}
 \frac{d\bar{x}}{\bar{v}_e(\bar{x})},
\end{equation}
where the velocity $\bar{v}_e(\bar{x})$ is found from the energy
integral
\begin{equation}\label{eL0:gam_e-1=}
  \gamma_e(\bar{x}) -1 = 2\beta_0\Pi_0\bar{T} \,
  \ln\frac{4\bar{T}+\bar{x}_r}{4\bar{T}+\bar{x}}
\end{equation}
for an electron moving in the static potential (\ref{eL0:bar_phi=}).
After some algebra we obtain
\begin{equation}\label{eL0:t_eq=}
 \bar{t}_{eq}(\bar{x}_r) = \frac{4\bar{T}+
 \bar{x}_r}{\sqrt{\gamma_0\bar{T}}}
 \int\limits_0^{\sqrt{\ln(1+\bar{x}_r/4\bar{T})}}
 \frac{1+2\beta_0\Pi_0\bar{T} z^2}{\sqrt{1+\beta_0\Pi_0\bar{T} z^2}}\,
 e^{-z^2}\, dz.
\end{equation}
In the asymptotic limit of $\bar{x}_r \gg 1$ Eq.~(\ref{eL0:t_eq=})
yields $\bar{t}_{eq}(\bar{x}_r) = \alpha_{eq} \bar{x}_r$, where the
numerical coefficient $\alpha_{eq}$ is a weak function of $\Pi_0$:
$\alpha_{eq} =\frac{1}{2}\sqrt{3\pi}$ for $\Pi_0 \ll 1$, and
$\alpha_{eq} =1$ for $\Pi_0 \gg 1$. This leads us to a conclusion that
asymptotically the ratio $\bar{x}_r(\bar{t})/\bar{t}$ should approach a
certain constant value $\alpha_r$ (or oscillate in a narrow range
around this value), which is fully confirmed by the TIAC simulations
for times $\bar{t} \leq 100$.

\begin{figure}[htb]
\includegraphics[width=75mm]{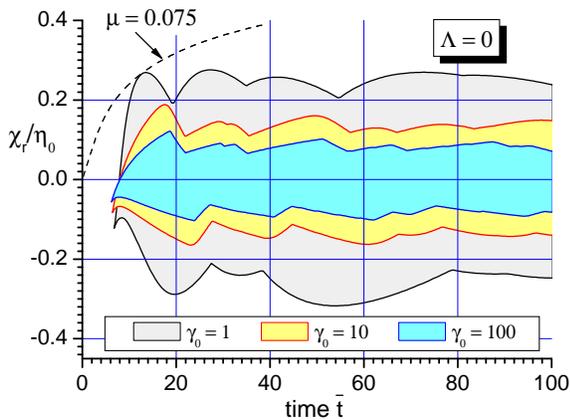}
\caption{Evolution of the relaxation zone (shaded areas) for $\Lambda
=0$ and three different values of $\gamma_0$ in terms of the hyperbolic
variable $\chi_r$ [see Eq.~(\ref{eL0:hi_r=})] normalized to $\eta_0 =
\textrm{arccosh}\gamma_0$ as calculated numerically with the TIAC code.
The upper and the lower branches of the $\chi_r(\bar{t})$ curves
correspond, respectively, to the boundaries in the upper ($x>0$) and
lower ($x<0$) half-spaces. Dashed curve shows the test ion trajectory
for $\gamma_0 =1$ and $\mu =0.075$.} \label{f:5}
\end{figure}

Figure~\ref{f:5} shows the temporal dependence of the upper and lower
boundaries of the relaxation zone in terms of a hyperbolic variable
$\chi_r = \chi_r(\bar{t})$ introduced via a relationship
\begin{equation}\label{eL0:hi_r=}
  \tanh\chi_r = \frac{x_r(t)}{ct} =\beta_0
  \frac{\bar{x}_r(\bar{t})}{\bar{t}}.
\end{equation}
The fraction of the electron cloud occupied by the upper laminar zone
is given by the difference $1-\chi_r/\eta_0$ [in the ultra-relativistic
case the separation along the hyperbolic variable $\chi =
\textrm{arctanh}(x/ct)$ is physically more representative than the
separation along $x$], where $\eta_0 =\textrm{arccosh}\,\gamma_0=
\textrm{arcsinh}\,\Pi_0$. If $\chi_r(\bar{t})$ should reach the value
$\eta_0$, it would mean that the relaxation zone has reached the
electron front $\bar{x} = \bar{t}$ and the laminar zone has vanished.
One clearly sees that typically the electron cloud is not dominated by
the relaxation zone, whose fraction asymptotically approaches some
constant value, and which has a tendency to shrink with the increasing
$\gamma_0$. For $\gamma_0 =1$ this fraction is $\alpha_r =0.25 \pm
0.02$. Therefore, it can be expected that there exists a sizable window
in the parameter space where the test ion acceleration takes place
either entirely or predominantly in the laminar zone.

\subsection{Solution for $\Lambda >0$} \label{s:eL}

For a finite-thickness foil it is convenient to introduce a different
pair of time and length units,
\begin{equation}\label{eL:[t]=}
  [t] =\Lambda t_{\Sigma} =\omega_0^{-1}\sqrt{\gamma_0}, \quad
  [x]= \Lambda l_{\Sigma} =v_0\omega_0^{-1}\sqrt{\gamma_0},
\end{equation}
where the time unit $[t]$ is based on the relativistic plasma frequency
$\omega_0\gamma_0^{-1/2}$. In these units the frequency and the
amplitude of the electron plasma oscillations are of the order unity
for any value of $\gamma_0 \geq 1$. Below, the quantities measured in
units (\ref{eL:[t]=}) are marked with a tilde. Note that in these units
the dimensionless thickness of the foil is $\tilde{l}_0 \equiv
\Lambda$.

\subsubsection{Electron trajectories}

For subsequent analysis of the test ion motion we need the electron
trajectories in the upper half-space $x> 0$. However, because for
$\Lambda >0$ each such trajectory $\tilde{x}_e(\tilde{t},\tilde{l})$
starts inside the foil at $-\Lambda \leq \tilde{x} \leq 0$, we should
solve Eqs.~(\ref{em:dx,p_e/dt=}) in this region as well, with the
initial conditions $\tilde{x}_e(0,\tilde{l}) = -\tilde{l}$,
$\Pi_e(0,\tilde{l}) =\Pi_0$. The required relativistic solution in the
laminar zone at $-\Lambda \leq \tilde{x}_e \leq 0$ has the form
\begin{equation}\label{eL:x_eh=}
  \tilde{x}_e(\tilde{t},\tilde{l},\beta_0) = -\tilde{l}
  +h(\tilde{t},\beta_0),
\end{equation}
where $h(\tilde{t},\beta_0)$ is a periodic function of time
$\tilde{t}$; for the first quarter-period it is implicitly given by the
quadrature
\begin{equation}\label{eL:h=}
  \tilde{t} = \int\limits_0^{h} \frac{1-\frac{1}{2} \beta_0^2
  \zeta^2}{\sqrt{1- \zeta^2 +\frac{1}{4} \beta_0^2
  \zeta^4}}\, d\zeta
\end{equation}
Actually, this is a solution for a relativistic particle moving in a
quadratic oscillator potential, for which one has the following energy
integral
\begin{equation}\label{eL:e-en-int=}
  \sqrt{1+\Pi_e^2} +\frac{1}{2} \beta_0 \Pi_0\, h^2 =\gamma_0.
\end{equation}
From Eq.~(\ref{eL:e-en-int=}) one readily establishes that the full
amplitude (in units of $v_0\omega_0^{-1} \sqrt{\gamma_0}$) of plasma
oscillations inside the foil is $2h_0$, where
\begin{equation}\label{eL:h_0=}
  h_0 = \left(\frac{2\gamma_0}{\gamma_0 +1}\right)^{1/2}.
\end{equation}
In the non-relativistic limit, obtained by putting $\beta_0 =0$ in
Eq.~(\ref{eL:h=}), we have $h(\tilde{t}) = \sin \tilde{t}$ and $h_0
=1$.

Now, the trajectory of an electron $\tilde{l}$ in the laminar zone at
$x>0$, which matches the solution (\ref{eL:x_eh=}), (\ref{eL:h=}) at a
point $\tilde{t} =\tilde{t}_{e1}$, $\tilde{x}_e= 0$, can be written as
\begin{eqnarray}\label{eL:tx_e=}
  \tilde{x}_e(\tilde{t},\tilde{l},\beta_0) &=& \frac{R}{P+
  \sqrt{P^2-\beta_0^2 \tilde{l}\, R}}, \\ \label{eL:Pi_e=}
  \Pi_e(\tilde{t},\tilde{l},\beta_0) &=& \Pi_0\left[Q- \tilde{l}\,
  (\tilde{t} -\tilde{t}_{e1})\right],
\\ \label{eL:P=}
  P=P(\tilde{l},\beta_0) &=& 1-\frac{1}{2}\, \beta_0^2\, \tilde{l}^2,
  \\ \label{Q=}
  Q=Q(\tilde{l},\beta_0) &=& \sqrt{1- \tilde{l}^2 +
  \frac{1}{4}\, \beta_0^2\, \tilde{l}^4},
  \\ \label{eL:R=}
  R=R(\tilde{t},\tilde{l},\beta_0) &=&  (\tilde{t} -\tilde{t}_{e1})
  \left[2Q -\tilde{l}\, (\tilde{t} -\tilde{t}_{e1})\right],
\\ \label{eL:tt_e1=}
  \tilde{t}_{e1}=\tilde{t}_{e1}(\tilde{l},\beta_0) &=&
  \int\limits_0^{\tilde{l}} \frac{1-\frac{1}{2}
  \beta_0^2 \zeta^2}{\sqrt{1- \zeta^2 +\frac{1}{4} \beta_0^2
  \zeta^4}}\, d\zeta;
\end{eqnarray}
here the values of $\tilde{l}$ are confined to the interval $0 \leq
\tilde{l} \leq \min\{\Lambda,h_0\}$. The non-relativistic limit of this
solution is easily recovered by putting $\beta_0 =0$. To obtain the
electric field in the laminar zone
\begin{equation}\label{eL:E=}
  E(\tilde{t},\tilde{x}) = 4\pi en_0 v_0\omega_0^{-1}\gamma_0^{1/2}\;
  \tilde{l}(\tilde{t},\tilde{x}),
\end{equation}
one needs the function $\tilde{l}(\tilde{t},\tilde{x},\beta_0)$, which
is the inverse of $\tilde{x}_e(\tilde{t},\tilde{l},\beta_0)$ with
respect to $\tilde{l}$. When calculating the results discussed in
section~\ref{s:iL}, such inversion of
Eqs.~(\ref{eL:tx_e=})-(\ref{eL:tt_e1=}) was performed numerically.

Since the Lagrangian coordinate $\tilde{l}$ belongs to the interval
$0\leq \tilde{l} \leq \Lambda$, a vacuum gap is formed between the
ejected electrons and the ion layer in the case of $\Lambda< h_0$ for a
limited time after the rear electron edge $\tilde{l} = \Lambda$ reaches
the foil surface at $x=0$ --- similar to the case of $\Lambda=0$ (see
Fig.~\ref{f:2}) where such a gap exists at $0< \bar{t}< 2$. Throughout
the gap, the electric field $E$ is constant and $\tilde{l}= \Lambda$.
To avoid unnecessary technical complications, we exclude the interval
$0< \Lambda< h_0< \sqrt{2}$ from our consideration, i.e.\ we perform
all the calculations either for the limiting case of $\Lambda=0$, or
for $\Lambda> h_0$ when no gap appears between the ejected electrons
and the foil ions. The reward is that we get rid of the dependence of
$\tilde{l}$ on the parameter $\Lambda$, i.e.\
$\tilde{l}(\tilde{t},\tilde{x},\beta_0)$ is a smooth function of
$\tilde{t}$ and $\tilde{x}$ for all $\tilde{t}\geq 0$ and $0\leq
\tilde{x} \leq\tilde{t}$, and this function does not depend on
$\Lambda$. In particular, this leads us to an important conclusion that
the behavior of the electron trajectories (hence, the distribution of
any other physical quantity) in the laminar zone of the electron cloud
at $\tilde{x}>0$ does not depend on the parameter $\Lambda$ for
$\Lambda> h_0$.

\subsubsection{Evolution of the relaxation zone} \label{s:rlxL}

\begin{figure}[htb]
\includegraphics[width=75mm]{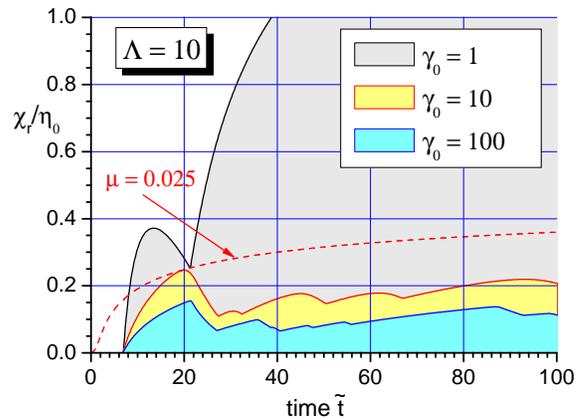}
\caption{Same as Fig.~\ref{f:5} but for a finite foil thickness
$\Lambda =10$. Shown here is the relaxation zone in the upper
half-space $x>0$ above the initial foil surface $x=0$. Dashed curve is
the test ion trajectory for $\gamma_0 =10$ and $\mu =0.025$.}
\label{f:6}
\end{figure}

The general arguments on the evolution of the relaxation zone
formulated in section~\ref{s:rlxL0} for $\Lambda=0$ remain valid at
$\Lambda> 0$ as well. In particular, these arguments lead us to the
following estimate for the relative width of the relaxation zone
\begin{equation} \label{eL:alfa_hir=}
\left|\frac{\chi_r(t)}{\eta_0}\right| \leq \alpha_{\chi r} < \infty,
\end{equation}
where $\chi_r(t)$ is defined in Eq.~(\ref{eL0:hi_r=}), and
$\alpha_{\chi r}$ does not depend on time $t$ but is a function of
$\Lambda$ and $\gamma_0$. Unfortunately, we have no rigorous proof that
one can introduce an upper bound (\ref{eL:alfa_hir=}) valid at all
times $t< \infty$. We have only been able to verify it numerically with
the TIAC code within a limited range of $\Lambda \lesssim 20$,
$\tilde{t} \lesssim 100$. As an example, Fig.~\ref{f:6} shows the
evolution of the relaxation zone for $\Lambda =10$ and $\gamma_0 =1$,
10, and 100.

The TIAC simulations clearly indicate that the relative width of the
relaxation zone $\alpha_{\chi r}$ increases with the increasing
$\Lambda$, and decreases with the increasing $\gamma_0$. In particular,
in the non-relativistic limit of $\gamma_0 \to 1$ we have $\alpha_{\chi
r}> 1$ for $\Lambda> 5.45$, i.e.\ for $\Lambda> 5.45$ the laminar zone
completely vanishes after some time (see Fig.~\ref{f:6}). However, it
reappears already at moderately relativistic electron energies
$\gamma_0 \gtrsim 4$, and becomes quite broad at $\gamma_0> 10$--100,
when we have $\alpha_{\chi r} \lesssim 0.2$--0.3. The latter implies
that the laminar-zone regime of test ion acceleration should be
particularly relevant to highly relativistic electron clouds with
$\gamma_0 \gtrsim 100$.

In the previous subsection it was established that the electron
trajectories in the laminar zone do not depend on $\Lambda$ for
$\Lambda> h_0$. This, however, does not mean that the same should apply
to the boundary  between the relaxation and the laminar zones because
this boundary is determined by electrons traversing the relaxation
zone. Nevertheless, one might expect that the curve $\chi_r(\tilde{t})$
should approach a certain limiting form for a fixed $\gamma_0$ and
$\Lambda \to \infty$: such a limit would correspond to the case where
the bulk plasma ions occupy the entire half-space $x<0$. However, the
existence of such limit for all $t< \infty$ is far from obvious.
Numerical simulations show that only the initial portion of the
$\chi_r(\tilde{t})$ curve around its first local maximum at $\tilde{t}
\simeq 15$--20 (see Figs.~\ref{f:5} and \ref{f:6}) becomes independent
of $\Lambda$ for $\Lambda \gtrsim 2.5$.

\section{Motion of test ions} \label{s:im}

Acceleration of a positive test ion with a charge $+eZ_p$ and a mass
$m_p$ (in conventional units) is described by the following equations
of motion
\begin{subequations} \label{im:dx,p_p/dt=}
\begin{eqnarray}\label{im:dx_p/dt=}
  \frac{1}{c}\frac{dx_p}{dt} &=&
  \frac{\Pi_p}{\sqrt{1+\Pi_p^2}} \equiv\beta_p, \\ \label{im:dPi_p/dt=}
  \frac{d\Pi_p}{dt} &=& +\frac{eZ_p}{m_pc} \, E(t,x_p),
\end{eqnarray}
\end{subequations}
where $x_p(t)$ is the position of the ion, and $\Pi_p(t)=
\beta_p\gamma_p =\beta_p(1-\beta_p^2)^{-1/2}$ is its momentum in units
of $m_pc$. As a rule, we assume that the test ion starts at $t=t_{p0}
=0$ with the initial values $x_p(t_{p0})= \Pi_p(t_{p0}) =0$. Then,
because the electric field $E(t,x)$ is non-negative for all $x\geq 0$,
we are guaranteed that $x_p(t)> 0$ for all $t>0$. Equations
(\ref{im:dx,p_p/dt=}) bring in a dimensionless parameter $\mu
=m_eZ_p/m_p$, which contains all the necessary information about the
test ion.

\subsection{Solution for $\Lambda =0$} \label{s:iL0}

Here, as in section~\ref{s:eL0}, we use the units (\ref{eL0:[t]=}) and
mark thus normalized quantities with a bar. The accelerating electric
field $E(\bar{t},\bar{x})$ in the laminar zone is given by
Eq.~(\ref{eL0:E=}).

\subsubsection{Vacuum phase}

When $\Lambda=0$, a test ion begins its motion by passing through a
vacuum gap, where it is accelerated by a constant field $E= 4\pi
e\Sigma$, until its trajectory
\begin{equation}\label{iL0:x_p-vac=}
  \bar{x}_{p,vac}(\bar{t}) =\frac{1}{\mu\beta_0\Pi_0} \left( \sqrt{1+
  \mu^2\Pi_0^2 \bar{t}^2} -1\right)
\end{equation}
crosses the rear edge [$\xi=1$ in Eq.~(\ref{eL0:bx_e-1=})] of the
electron cloud at
\begin{equation}\label{iL0:t_p1=}
  \bar{t}= \bar{t}_{p1} = \frac{2(1+\mu\gamma_0)}{1+2\mu\gamma_0+\mu^2},
  \quad \bar{x} =\bar{x}_{p1} = \bar{x}_{p,vac}(\bar{t}_{p1})
\end{equation}
with a momentum $\Pi_{p1}= \mu\Pi_0\bar{t}_{p1}$. The latter values
should be used as the initial conditions for further acceleration
inside the electron cloud.

\subsubsection{Non-relativistic solution}

In the non-relativistic limit a general analytical solution is easily
found to Eqs.~(\ref{im:dx,p_p/dt=}). The type of this solution depends
on whether the roots $\lambda$ and $1-\lambda$ of the characteristic
equation
\begin{equation}\label{iL0:char_eq=}
  \lambda^2 -\lambda +2\mu =0
\end{equation}
are real or complex. In this way a critical value $\mu_{\ast}
=\frac{1}{8}$ of the parameter $\mu$ is established, which separates
the two solution types. For $\mu> \frac{1}{8}$, when $\lambda$ is
complex, a test ion always catches up with the electron front $\bar{x}
=\bar{t}$ and acquires the final velocity $v_{p\infty}$ in excess of
the initial electron velocity $v_0$.

In the physically more interesting case of $\mu< \frac{1}{8}$ (a
sufficiently heavy test ion) the sought for solution of
Eqs.~(\ref{im:dx,p_p/dt=}) is given by
\begin{equation}\label{iL0:nr-x_p=}
  \bar{x}_p(\bar{t}) = \bar{t} -C_1\bar{t}^{1-\lambda}
  +C_2\bar{t}^{\lambda},
\end{equation}
where
\begin{equation}\label{iL0:nr-lam=}
  \lambda = \frac{1}{2}\left(1-\sqrt{1-8\mu}\right)
\end{equation}
is the smaller of the two real roots of Eq.~(\ref{iL0:char_eq=}), and
the integration constants
\begin{equation}\label{iL0:C_1,2=}
  C_1= \frac{2^{\lambda}(1-\lambda-\mu)}{(1- 2\lambda)(1+
  \mu)^{1+\lambda}}, \quad C_2= \frac{2^{1-\lambda}(\lambda-
  \mu)}{(1- 2\lambda)(1+ \mu)^{2-\lambda}}
\end{equation}
are calculated by using the initial conditions (\ref{iL0:t_p1=}) for
$\gamma_0 =1$; Eq.~(\ref{iL0:nr-x_p=}) applies at $\bar{t} \geq
\bar{t}_{p1}= 2/(1+\mu)$.

Solution (\ref{iL0:nr-x_p=}) reveals the following general features of
ion acceleration in the laminar zone. The distance $\bar{t}
-\bar{x}_p(\bar{t})$ between the electron front and the accelerated ion
increases monotonically with time, i.e.\ a test ion with $\mu<
\frac{1}{8}$ lags further and further behind the electron front as
$t\to\infty$. At the same time, the ion velocity $v_p(t) =dx_p/dt$
monotonically grows in time, and in the formal limit of $t\to\infty$ it
asymptotically approaches the initial electron velocity $v_0$ --- as it
has been established earlier in Ref.~\cite{BuEs.04}. The latter,
however, occurs on an extremely long timescale $\bar{t}_{ac} \simeq
\exp\left(\frac{1}{2}\mu^{-1}\right)$, which is beyond any realistic
value for protons and other ions with $\mu \leq 1/1836$. From practical
point of view, an intermediate asymptotics
\begin{equation}\label{iL0:nr-v_p_as=}
  v_p(t) \approx 2\mu v_0 \,
  \ln\left(\frac{2\pi e^2\Sigma}{m_ev_0}\, t\right),
\end{equation}
inferred from Eqs.~(\ref{iL0:nr-x_p=})--(\ref{iL0:C_1,2=}) for $\mu \ll
1$ and $1 \ll \bar{t} \ll \exp\left(\frac{1}{2}\mu^{-1}\right)$, might
be of interest --- if not the presence of the relaxation zone. The fact
is that solution (\ref{iL0:nr-x_p=}) applies only for ions with $0.0745
\leq \mu < 0.125$ because, as one finds from the TIAC simulations (see
Fig.~\ref{f:5}), the ion trajectories for $\Lambda=0$ and $\mu< 0.0745$
penetrate into the relaxation zone. Ions with $\mu \leq 1/1836$ are
accelerated deep in the relaxation zone, where one can use the
Boltzmann relation and the quasi-static potential (\ref{eL0:bar_phi=});
one readily verifies that the potential (\ref{eL0:bar_phi=}) leads to
much higher (by roughly a factor $\mu^{-1}$) final ion energies than
those obtained from Eq.~(\ref{iL0:nr-v_p_as=}). From this we conclude
that the acceleration in the near-front laminar zone of a
non-relativistic electron cloud is never important for protons and
heavier test ions. In section~\ref{s:iL} we demonstrate that this
conclusion, proved here for $\Lambda=0$, is valid for $\Lambda>1$ as
well.

\subsubsection{Relativistic solution}

Analysis of ion motion in the general relativistic case is
significantly simplified after we make a transformation from the
dynamic variables $x_p$ and $\Pi_p$ to hyperbolic variables $\chi$ and
$\eta$ defined by means of the relationships
\begin{equation}\label{iL0:hi=}
  \tanh  \chi = \frac{x_p}{ct} =\beta_0\frac{\bar{x}_p}{\bar{t}}, \quad
  \sinh\eta = \Pi_p.
\end{equation}
Accordingly, the initial electron velocity $v_0$ is represented by a
parameter $\eta_0$, where $\Pi_0=\sinh\eta_0$, $\gamma_0 =\cosh\eta_0$,
$\beta_0 =\tanh\eta_0$. In terms of these variables
equations~(\ref{im:dx,p_p/dt=}) with the expression (\ref{eL0:E=}) for
the electric field become
\begin{subequations}\label{iL0:dhi,eta/dzeta=}
\begin{eqnarray}\label{iL0:dhi/dzeta=}
  \frac{d\chi}{d\zeta} &=& \frac{\cosh\chi}{\cosh\eta}\, \sinh(\eta-\chi)
  \\ \label{iL0:deta/dzeta=}
  \frac{d\eta}{d\zeta} &=& 2\mu\, \frac{\cosh\chi}{\cosh\eta}\,
  \sinh(\eta_0-\chi),
\end{eqnarray}
\end{subequations}
where $\zeta = \ln\bar{t}$. Equations (\ref{iL0:dhi,eta/dzeta=}) do not
contain the independent variable $\zeta$ on their right-hand sides,
which means that certain key features of the ion motion can be analyzed
by inspecting the integral curves of the first-order phase equation
\begin{equation}\label{iL0:deta/dhi=}
  \frac{d\eta}{d\chi} = 2\mu\, \frac{\sinh(\eta_0
  -\chi)}{\sinh(\eta-\chi)}
\end{equation}
in the ($\chi, \eta$) plane. Of principal importance here is the
singular point $(\chi,\eta) =(\eta_0,\eta_0)$.

\begin{figure}[htb]
\includegraphics[width=75mm]{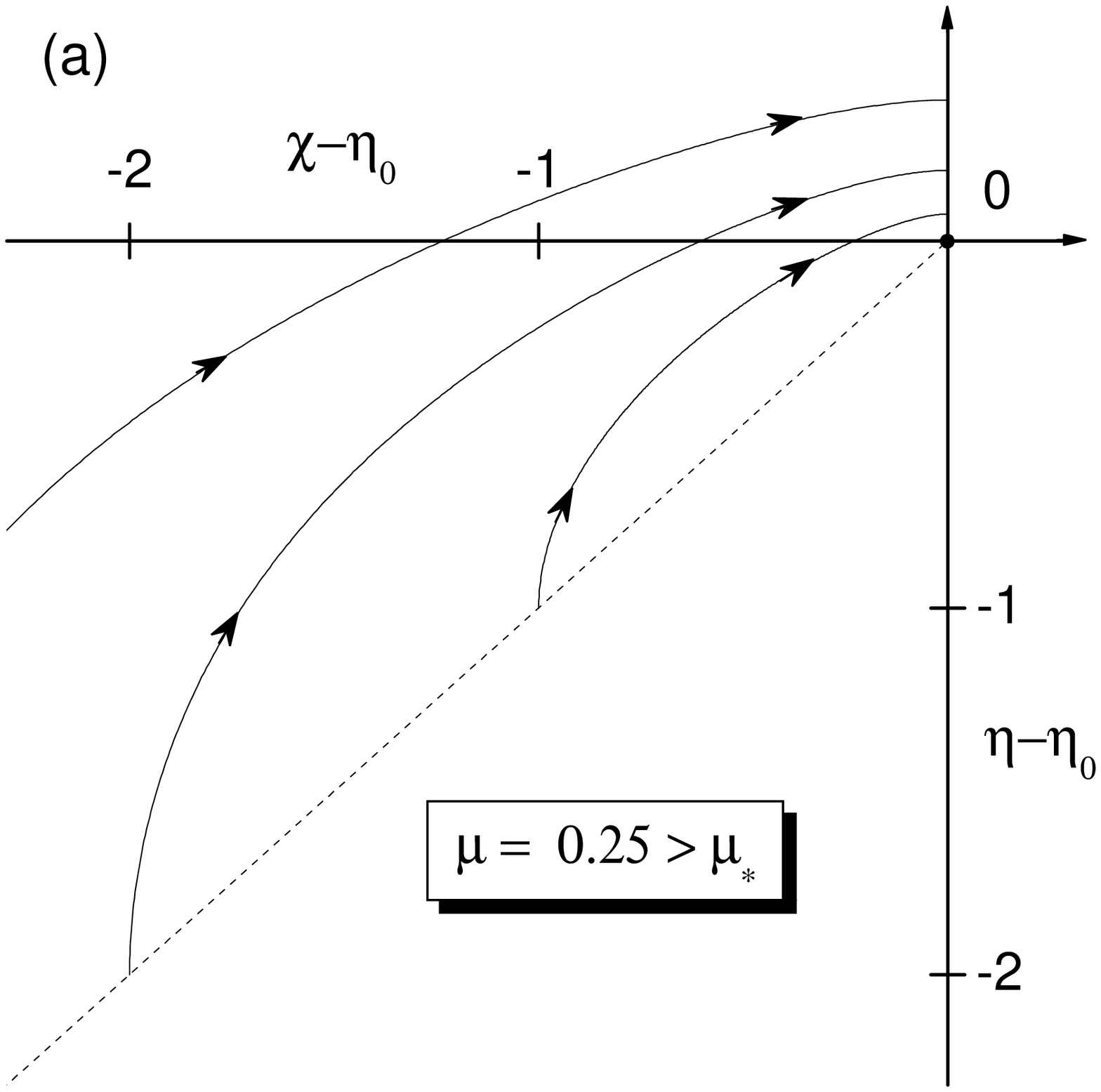}

\vspace{5mm}

\includegraphics[width=75mm]{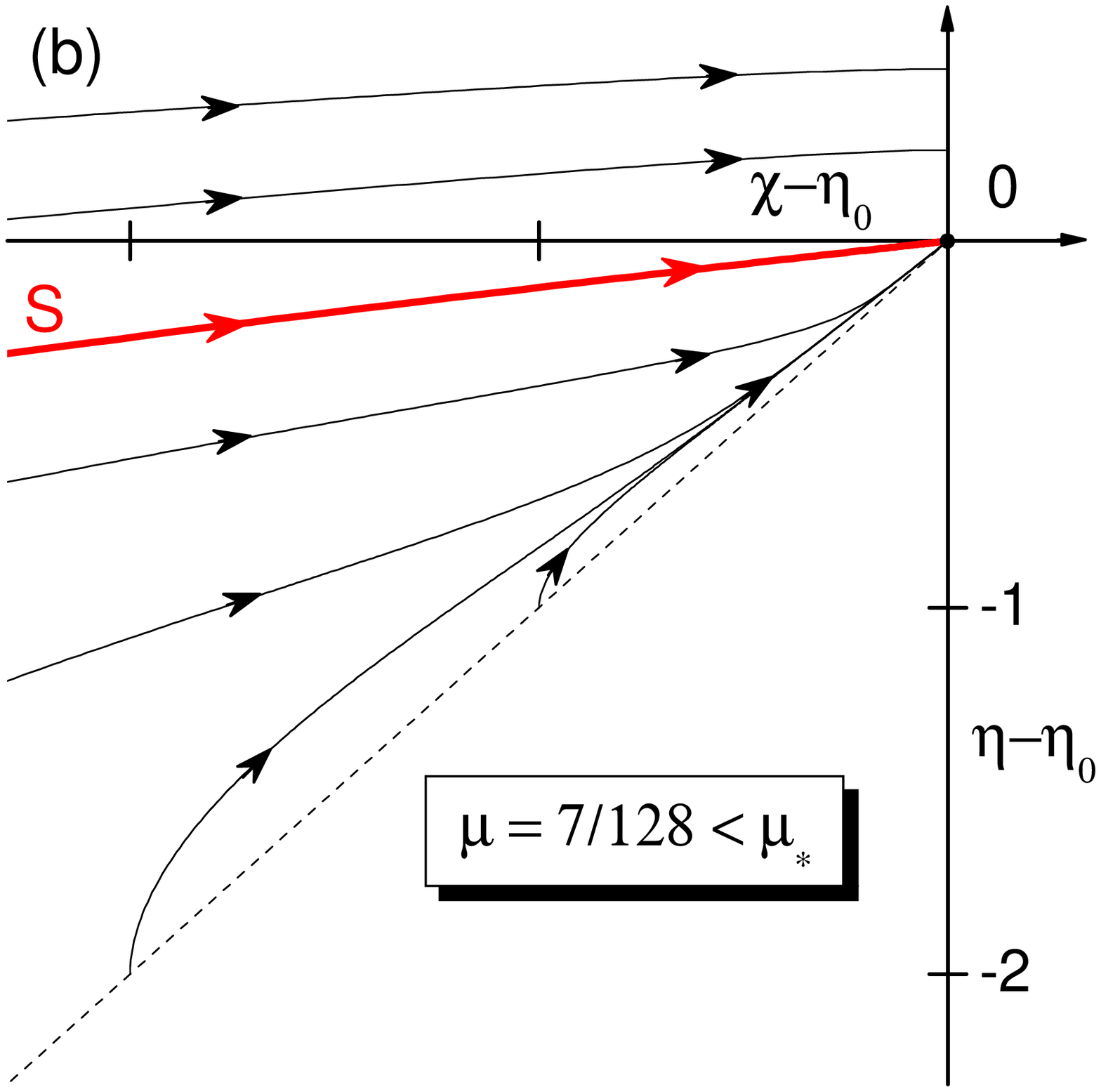}
\caption{Phase trajectories for the ion equations of motion in the form
(\ref{iL0:dhi,eta/dzeta=}) in the vicinity of the singular point
$(\chi,\eta) =(\eta_0,\eta_0)$ for two values of parameter $\mu$. The
singular point is a focus for $\mu> \mu_{\ast} = \frac{1}{8}$ (a), and
a node for $\mu< \mu_{\ast}$ (b). The separatrix~$S$ is a solitary
integral curve of Eq.~(\ref{iL0:deta/dhi=}) which enters the node
$(\eta_0,\eta_0)$ along the direction $\eta-\eta_0 =
\lambda(\chi-\eta_0)$, where $\lambda$ is given by
Eq.~(\ref{iL0:nr-lam=}).} \label{f:7}
\end{figure}

First of all note that physically meaningful in our context are the
integral curves of Eq.~(\ref{iL0:deta/dhi=}) that lie in the half-plane
$\eta> \chi$: this follows from inequality $x_p(t)< v_pt$ valid for any
motion with $x_p(0)\geq 0$ and a monotonically increasing velocity
$v_p(t)$. The electron front $x = v_0t$ is represented by the vertical
line $\chi =\eta_0$. A test ion can cross the electron front either in
a regular way at $\eta> \eta_0$, or by passing through the singular
point $(\eta_0,\eta_0)$ along one of the two characteristic directions
defined by the characteristic equation (\ref{iL0:char_eq=}) (see
Fig.~\ref{f:7}). A remarkable fact is that the roots of
Eq.~(\ref{iL0:char_eq=}) do not depend on $\eta_0$, i.e.\ are the same
for the relativistic and non-relativistic motions. As a consequence, we
obtain a universal critical value $\mu_{\ast} =\frac{1}{8}$ which
separates two topologically different patterns of the ion trajectories
near the singular point $(\eta_0,\eta_0)$.

For light ions with $\mu> \frac{1}{8}$, when the singular point
$(\eta_0,\eta_0)$ is a focus (see Fig.~\ref{f:7}a), the qualitative
picture is the same as in the non-relativistic case: a test ion always
reaches the electron front $\chi=\eta_0$ within a finite time interval
and crosses it at $\eta> \eta_0$, i.e.\ with a velocity
$v_p=v_{p\infty} > v_0$.

\begin{figure}[htb]
\includegraphics[width=75mm]{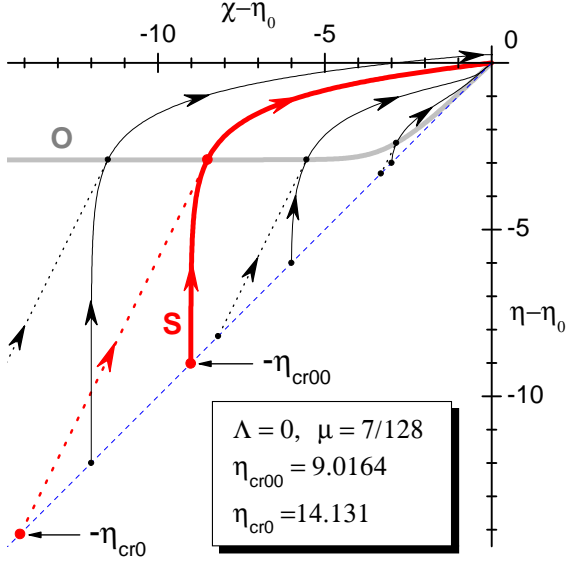}
\caption{Global view of the relativistic phase trajectories of a test
ion with $\mu =7/128$. Depending on the ion start delay $\bar{t}_{p0}
\geq 0$, the starting point of each trajectory may lie anywhere between
its intersection with the grey thick curve~$O$, defined by
Eqs.~(\ref{iL0:hi,eta_p1=}), and with the bisector $\eta=\chi$ (dashed
line). Straight dotted segments of the phase trajectories correspond to
ion acceleration inside the vacuum gap.} \label{f:8}
\end{figure}

For heavier ions with $\mu< \frac{1}{8}$ the singular point
$(\chi,\eta)= (\eta_0,\eta_0)$ is a node with two entrance directions
\begin{subequations}\label{iL0:r-dir=}
\begin{eqnarray}\label{iL0:r-dir_a=}
  \!\!\! \eta-\eta_0  =  \lambda (\chi-\eta_0),  &&\mbox{separatrix $S$},
  \\ \label{iL0:r-dir_b=}  \!\!\! \eta-\eta_0  =
  (1-\lambda)(\chi-\eta_0),  && \mbox{general direction},
\end{eqnarray}
\end{subequations}
where $\lambda$ is given by Eq.~(\ref{iL0:nr-lam=}). Note that for
$\mu\ll 1$ we have $\lambda \approx 2\mu \ll 1$. The separatrix~$S$
divides all the integral curves in the $(\chi,\eta)$ plane into two
classes. To the first class belong the curves which lie below the
separatrix~$S$ in Figs.~\ref{f:7}b and \ref{f:8} and enter the singular
point along the general direction (\ref{iL0:r-dir_b=}). From
Eqs.~(\ref{iL0:dhi,eta/dzeta=}), (\ref{iL0:deta/dhi=}) and
(\ref{iL0:r-dir_b=}) one readily verifies that these curves approach
the electron front $\chi =\eta_0$  in the asymptotic limit of $\bar{t}
\to \infty$, with the value of $\eta_0 -\chi$ falling off as
$\bar{t}^{-\lambda}$. Exactly as in the non-relativistic limit, the
distance $\bar{t}- \bar{x}_p(\bar{t})$ to the electron front increases
monotonically in direct proportion to $\bar{t}^{1-\lambda}$ as $\bar{t}
\to \infty$, while the ion momentum $\Pi_p(\bar{t})$ monotonically
grows in time and asymptotically approaches $\Pi_0$ on a timescale
$\bar{t}_{ac} \simeq \exp\left(\lambda^{-1}\right)$. To the second
class belong the trajectories that lie above the separatrix~$S$ and
cross the electron front $\chi = \eta_0$ at $\eta> \eta_0$. Ions moving
along such trajectories overtake the electron front within a finite
time and reach the final velocity $v_{p\infty}> v_0$
($\gamma_{p\infty}> \gamma_0$).

A qualitative difference between the relativistic and non-relativistic
cases arises when one considers the behavior of the separatrix $S$. In
the non-relativistic limit the latter is a straight line given by
Eq.~(\ref{iL0:r-dir_a=}). Therefore, all physically interesting ion
trajectories that start from $x_p=0$ with the zero initial velocity
$v_p=0$  at any time $t_{p0} \geq 0$ always lie below $S$, i.e.\ belong
to the first class described in the previous paragraph. The
relativistic separatrix, in contrast, bends down and crosses the line
$\eta=\chi$ at a certain value $\eta- \eta_0= \chi- \eta_0
=-\eta_{cr00}$ (see Fig.~\ref{f:8}), where $\eta_{cr00}
=\eta_{cr00}(\mu)$ is a function of $\mu$. As a consequence, for a
given $\mu$ and $\eta_0> \eta_{cr00}(\mu)$ the phase trajectory of a
test ion may pass above the separatrix and fall into the second class.
In such a case a test ion overtakes the electron front within a finite
time interval. To reach more definite conclusions, we have to take a
closer look at the initial conditions.

If a test ion begins to move simultaneously with electrons at $t=
t_{p0} =0$, the initial part of its trajectory lies in vacuum and is
represented by a segment of a straight line $\eta= \eta_{vac}(\chi) =
2\chi$ [as it follows from Eqs.~(\ref{iL0:x_p-vac=}) and
(\ref{iL0:hi=})] with $0\leq \chi \leq \chi_{p1} =\frac{1}{2}
\eta_{p1}$; in Fig.~\ref{f:8} these segments are shown as dotted
straight intervals. Then, the initial conditions for the phase equation
(\ref{iL0:deta/dhi=}) are given by the values
\begin{subequations} \label{iL0:hi,eta_p1=}
\begin{eqnarray} \label{iL0:hi_p1=}
\hspace{-8mm}
  &&\chi = \chi_{p1} = \frac{1}{2} \eta_{p1},
  \\ \label{iL0:eta_p1=}
\hspace{-8mm}
  &&\eta = \eta_{p1} = \textrm{arcsinh}\left[ \frac{2\mu\sinh\eta_0 \,
  \left(1+ \mu \cosh\eta_0\right)}{1+2\mu \cosh\eta_0 +\mu^2}\right]\!,
\end{eqnarray}
\end{subequations}
inferred from Eqs.~(\ref{iL0:x_p-vac=}), (\ref{iL0:t_p1=}) and
(\ref{iL0:hi=}). If we fix $\mu$ and treat $\eta_0$ as a free
parameter, Eqs.~(\ref{iL0:hi,eta_p1=}) define  a universal curve~$O$,
the locus of the initial points for the integral curves of
Eq.~(\ref{iL0:deta/dhi=}) in the $(\chi-\eta_0,\eta-\eta_0)$ plane (see
Fig.~\ref{f:8}). The value of parameter $\eta_0$ along the curve~$O$ at
its intersection with the separatrix~$S$ defines the primary critical
value $\eta_{cr0} =\eta_{cr0}(\mu)$ for this parameter. Its meaning is
as follows: for any $\eta_0> \eta_{cr0}(\mu)$ a test ion with the
charge-over-mass ratio $\mu$ finally catches up with the electron front
and overtakes it.

The fact that the vacuum segments of the phase trajectories in
Fig.~\ref{f:8} rise less steeply than the initial portions of the
relativistic integral curves of Eq.~(\ref{iL0:deta/dhi=}) implies that
a delayed (at $t_{p0}>0$) start of a test ion may result in its more
efficient acceleration. In reality such a delayed start may occur when
a test positive particle is created on the spot some time after the
laser pulse. If we consider a delayed start at $\bar{t}_{p0} \geq 2$,
when the vacuum gap is already closed (the opposite extreme to the
previously considered case of simultaneous start at $t_{p0} =0$), the
locus of the initial points $(\chi,\eta)= (0,0)$ for the integral
curves of Eq.~(\ref{iL0:deta/dhi=}) in the $(\chi-\eta_0,\eta-\eta_0)$
plane  will be the bisector line $\eta-\eta_0=\chi-\eta_0$. Hence, the
intersection of the separatrix~$S$ with this bisector defines the
secondary critical value $\eta_{cr00} =\eta_{cr00}(\mu)<
\eta_{cr0}(\mu)$ of parameter $\eta_0$ (see Fig.~\ref{f:8}) which has
the following meaning: for any $\eta_0< \eta_{cr00}(\mu)$ a test ion
with the charge-over-mass ratio $\mu$ always stays behind the electron
front for all possible starting times $t_{p0} \geq 0$ . In the
intermediate case of $\eta_{cr00}< \eta_0< \eta_{cr0}$ a test ion can
either overtake the electron front or stay behind it, depending on the
start delay $0< \bar{t}_{p0} <2$. A selection of $\eta_{cr0}(\mu)$ and
$\eta_{cr00}(\mu)$ values calculated by solving
Eq.~(\ref{iL0:deta/dhi=}) numerically is given in Table~\ref{t:1}.

\renewcommand{\arraystretch}{1}
\begin{table}[hbt] \begin{center}
\caption{Critical parameters $\eta_{cr0}(\mu)$, $\eta_{cr00}(\mu)$,
and $\eta_{cr\Lambda}(\mu)$ as calculated for a selection of $\mu$
values by numerically integrating the ion equations of motion
(\ref{iL0:dhi,eta/dzeta=}).} \label{t:1} \vspace{1mm}
\begin{tabular}{@{\extracolsep{2mm}}r|r|r|r}
\multicolumn{1}{c|}{$1/\mu$} &  \multicolumn{1}{c|}{$\eta_{cr0}$} &
\multicolumn{1}{c|}{$\eta_{cr00}$} &
\multicolumn{1}{c}{$\eta_{cr\Lambda}$} \\ \hline
10  & 5.94216 & 4.51193 & 5.35497 \\
20  & 15.7848 & 9.88926 & 15.0917 \\
100 & 94.3602 & 49.9827 & 93.6670 \\
200 & 193.685 & 99.9916 & 192.990 \\
1000& 992.089 & 499.998 & 991.395 \\
2000& 1991.40 & 999.999 & 1990.71
\end{tabular}\end{center}\end{table}
\renewcommand{\arraystretch}{1}

\subsubsection{The ultra-relativistic limit}

In the physically important limit of $\mu \ll 1$ the functions
$\eta_{cr0}(\mu)$ and $\eta_{cr00}(\mu)$ can be calculated analytically
by using the ultra-relativistic ($\mu\Pi_0 \gg 1$) expansion of
Eqs.~(\ref{iL0:hi,eta_p1=}) for the curve~$O$ in Fig.~\ref{f:8},
\begin{equation}\label{iL0:ur-O=}
  \eta -\eta_0 = \ln\mu,
\end{equation}
and the integral
\begin{equation}\label{iL0:ur-integ=}
  \exp\left(\eta-\eta_0 \right) -2\mu (\chi -\eta_0) =C
\end{equation}
of Eq.~(\ref{iL0:deta/dhi=}) in the limit of $\eta_0-\chi \gg 1$,
$\eta-\chi \gg 1$, when the right-hand side of
Eq.~(\ref{iL0:deta/dhi=}) can be approximated as
$2\mu\exp(\eta_0-\eta)$. Having set the integration constant $C=1$, we
obtain the equation of the separatrix~$S$ in the
$(\chi-\eta_0,\eta-\eta_0)$ plane. Note that, although derived in the
limit of $\eta_0-\chi \gg 1$, this equation has a correct limiting
behavior at $\chi\to \eta_0$ as well. After we calculate the
intersection points of the separatrix $S$ with the curve~$O$ [as given
by Eq.~(\ref{iL0:ur-O=})] and with the bisector $\chi=\eta$, we find
\begin{eqnarray}\label{iL0:eta_cr=}
  \eta_{cr0} = \ln\mu+\frac{1}{\mu} -1,  && \gamma_{cr0} =\frac{1}{2}
  \,\mu\, \exp\left(\frac{1}{\mu} -1\right), \\ \label{iL0:eta_cr0=}
  \eta_{cr00} = \frac{1}{2\mu},  && \gamma_{cr00} =\frac{1}{2}
  \exp\left(\frac{1}{2\mu}\right).
\end{eqnarray}
The corresponding critical values $\gamma_{cr0}(\mu)$ and
$\gamma_{cr00}(\mu)$ of the parameter $\gamma_0$ are obtained by
applying the ultra-relativistic formula $\gamma = \cosh\eta \approx
\frac{1}{2} \exp\eta$. Comparison with the numerical results from
Table~\ref{t:1} shows that for $\mu^{-1}> 20$ the asymptotic formulae
(\ref{iL0:eta_cr=}), (\ref{iL0:eta_cr0=}) for $\eta_{cr0}$ and
$\eta_{cr00}$ are accurate to within 1.4\%.

Making use of the integral (\ref{iL0:ur-integ=}) with the values of $C
= \exp(\eta_{\infty}-\eta_0)> 1$, we calculate the limiting value
$\gamma_{p\infty} = \frac{1}{2}\exp\eta_{\infty}$ of the ion
gamma-factor in the case when the ion overtakes the electron front,
\renewcommand{\arraystretch}{1.6}
\begin{equation}\label{iL0:ur-gam_inf=}
  \gamma_{p\infty}= \left\{ \begin{array}{ll}
  \!\mu \gamma_0\left(1+\displaystyle\ln
  \frac{2\gamma_0}{\mu}\right), & \textrm{$\bar{t}_{p0}=0$ and
  $\gamma_0 > \gamma_{cr0}(\mu)$}, \\
  \! 2\mu\gamma_0\, \ln(2\gamma_0), & \textrm{$\bar{t}_{p0}
  \geq 2$ and $\gamma_0 > \gamma_{cr00}(\mu)$}.
  \end{array} \right.
\end{equation}
\renewcommand{\arraystretch}{1}
This our result for $\gamma_{p\infty}$ differs significantly from the
value $\gamma_{p\infty} = 2\gamma_0^2$ calculated earlier in Eq.~(35)
of Ref.~\cite{BuEs.04}, which we believe to be erroneous. It should be
noted, however, that Eq.~(\ref{iL0:ur-gam_inf=}) can hardly be of any
practical interest for protons and heavier ions because for $\mu \leq
1/1836$ the corresponding values of $\gamma_{cr0} \geq 2.3\times
10^{793}$ and $\gamma_{cr00} \geq 2.4\times 10^{398} $ are way too
large to be ever encountered in nature.

\subsubsection{Intermediate asymptotics for the ion energy}

Having established that in reality test ions of common interest, i.e.\
those with $\mu \leq 1/1836$, always stay behind the electron front,
and that their dimensionless momentum $\Pi_p(\bar{t})$  approaches the
electron value $\Pi_0$ on an unrealistically long timescale
$\bar{t}_{ac} \simeq \exp\left(\frac{1}{2}\mu^{-1}\right)$, a natural
step would be to look for an intermediate asymptotics for
$\Pi_p(\bar{t})$, valid at $1\ll \bar{t} \ll
\exp\left(\frac{1}{2}\mu^{-1}\right)$, that might be of practical
interest for the problem considered.

Once we let $\mu \ll 1$ and agree that $\gamma_0 \ll
\gamma_{cr00}(\mu)< \gamma_{cr0}(\mu)$, we can integrate
Eq.~(\ref{iL0:deta/dzeta=}) in the limit of $\bar{t} \gg 1$ by making
an approximation $\eta=\chi$, which is valid to the first order in
$\mu$ along the direction of general approach (\ref{iL0:r-dir_b=}) to
the singular point $(\chi,\eta)= (\eta_0,\eta_0)$. With the initial
condition
\begin{equation}\label{iL0:Ias-eta_p1=}
  \eta(\zeta_{p1}) = \eta_{p1}, \quad \sinh\eta_{p1} =\Pi_{p1}=
  \mu\Pi_0 \bar{t}_{p1},
\end{equation}
where $\zeta_{p1}= \ln\bar{t}_{p1}$ and $\bar{t}_{p1}$ is given by
Eq.~(\ref{iL0:t_p1=}), the result of this integration reads
\begin{equation}\label{iL0:Ias-int=}
  \tanh\frac{\eta-\eta_0}{2} = \tanh\frac{\eta_{p1}
  -\eta_0}{2} \, \exp\left[-2\mu(\zeta-\zeta_{p1})\right].
\end{equation}
Performing Taylor expansion of Eq.~(\ref{iL0:Ias-int=}) with respect to
the small parameter $0< 2\mu(\zeta-\zeta_{p1}) \ll 1$, we derive the
following asymptotic expression for the test ion momentum $\Pi_p(t)$
\begin{equation}\label{iL0:Ias-Pi_p=}
  \frac{\gamma_p+\Pi_p}{\gamma_{p1}+\Pi_{p1}} =1+\left(
  \frac{\gamma_0+\Pi_0}{\gamma_{p1}+\Pi_{p1}} -
  \frac{\gamma_{p1}+\Pi_{p1}}{\gamma_0+\Pi_0}\right)\,
  \mu\ln\frac{t}{t_{p1}},
\end{equation}
where $\gamma_p =\left(\Pi_p^2+1\right)^{1/2}$ and  $\gamma_{p1}
=\left(\Pi_{p1}^2+1\right)^{1/2}$. Note that, once $1\ll \bar{t} \ll
\exp\left(\frac{1}{2}\mu^{-1}\right)$, Eq.~(\ref{iL0:Ias-Pi_p=})
applies at any degree of relativism of either electrons or a test ion,
i.e.\ any of the three quantities $\Pi_0$, $\Pi_{p1}$, and  $\Pi_{p}$
is allowed to be arbitrarily small or large compared to unity.
Comparison with numerical integration of
Eqs.~(\ref{iL0:dhi,eta/dzeta=}) shows that for $\mu \leq 1/1836$ the
error of the intermediate asymptotics (\ref{iL0:Ias-Pi_p=}) at $1\leq
\ln(t/t_{p1}) \leq 50$ is typically about 2--4\%, and never exceeds
8\%. In both the non-relativistic ($\Pi_0 \ll 1$) and the
ultra-relativistic ($\mu\Pi_0 \gg 1$) limits Eq.~(\ref{iL0:Ias-Pi_p=})
reduces to a simple expression
\begin{equation}\label{iL0:Ias-Pi_p==}
  \Pi_p \approx \Pi_{p1}
  \left(1+\ln\frac{t}{t_{p1}}\right).
\end{equation}

\subsection{Solution for $\Lambda> h_0$} \label{s:iL}

From practical point of view the case of $\Lambda \gtrsim 1$ is
generally more important than the limit $\Lambda=0$ considered so far.
As a typical example, it may be noted that a 1~${}\mu$m foil of solid
gold ionized to $z=50$ with $\gamma_0 =100$ would correspond to
$\Lambda \approx 10$. Here we prove that practically all the
qualitative and many of the quantitative results obtained for $\Lambda
=0$ extend to the case of $\Lambda> 0$ as well. To avoid unnecessary
mathematical complications, we exclude from our consideration the
intermediate range of $0< \Lambda< h_0$ and assume that $\Lambda> h_0 =
\left[ 2\gamma_0/(1 +\gamma_0)\right]^{1/2}$.

In units (\ref{eL:[t]=}) the equations of ion motion
(\ref{im:dx,p_p/dt=}) become
\begin{subequations} \label{iL:dx_p,Pi_p=}
\begin{eqnarray}\label{iL:dx_p=}
  \beta_0 \frac{d\tilde{x}_p}{d\tilde{t}} &=&
  \frac{\Pi_p}{\sqrt{1+\Pi_p^2}},
  \\ \label{iL:dPi_p=}
  \frac{d\Pi_p}{d\tilde{t}} &=& \mu \Pi_0\,
  \tilde{l}(\tilde{t},\tilde{x}_p,\beta_0).
\end{eqnarray}
\end{subequations}
As discussed in section~\ref{s:eL}, the function
$\tilde{l}(\tilde{t},\tilde{x},\beta_0)$ does not depend on $\Lambda$
at $\Lambda> h_0$, and is a smooth function of its arguments in the
entire region $0\leq \tilde{t} < \infty$, $0\leq \tilde{x} \leq
\tilde{t}$. In the case of a simultaneous ion start at $t_{p0}=0$ we
have the initial conditions $\tilde{x}_p(0)=\Pi_p(0) =0$, and the
leading terms in expansion of the desired solution to
Eqs.~(\ref{iL:dx_p,Pi_p=}) near $\tilde{t}= 0$ are given by
\begin{subequations}\label{iL:x_p,Pi_p(0)=}
\begin{eqnarray}\label{iL:x_p(0)=}
  \tilde{x}_p(\tilde{t}) &=& \frac{1}{6} \mu\gamma_0\, \tilde{t}^3 +
  O(\tilde{t}^4),
  \\ \label{iL:Pi_p(0)=}
  \Pi_p(\tilde{t}) &=& \frac{1}{2} \mu\Pi_0\, \tilde{t}^2
  +O(\tilde{t}^3).
\end{eqnarray}
\end{subequations}
Since neither Eqs.~(\ref{iL:dx_p,Pi_p=}) nor the pertinent boundary
conditions depend on $\Lambda$, we arrive at an important conclusion
that, when expressed in units (\ref{eL:[t]=}), all the results
concerning test ion acceleration in the laminar zone are independent of
$\Lambda$ for $\Lambda> h_0$.

The next important point is that in the limit of $\tilde{t}\to \infty$
equations~(\ref{iL:dx_p,Pi_p=}) become exactly equivalent to the
corresponding equations of motion for $\bar{x}_p(\bar{t})$,
$\Pi_p(\bar{t})$ in the case of $\Lambda=0$. To verify this, we note
that Eqs.~(\ref{eL:tx_e=})-(\ref{eL:tt_e1=}) imply  $\tilde{l} \sim
2/\tilde{t} \ll 1$ for $\tilde{t} \gg 1$. Then, by expanding
Eqs.~(\ref{eL:tx_e=})-(\ref{eL:tt_e1=}) in powers of $\tilde{l}$, we
derive an explicit formula
\begin{eqnarray}
\hspace{-7mm} \tilde{l}(\tilde{t},\tilde{x},\beta_0)  &\approx &
4(\tilde{t}-\tilde{x}) \left\{\tilde{t}^2-\beta_0^2\tilde{x}^2+2+
\vphantom{\left[(\tilde{t}^2-2)\right]^{1/2}}  \right.  \nonumber
\\ \label{iL:l_as1=} && \left.
\hspace{-3mm} \left[(\tilde{t}^2- \beta_0^2 \tilde{x}^2 -2)^2 +
8(1-\beta_0^2)\tilde{t}\tilde{x}\right]^{1/2}\right\}^{-1} \!\!\!\! ,
\end{eqnarray}
which is valid for any $0< \beta_0 <1$ in the limit of $\tilde{l} \ll
1$, and which is further simplified to
\begin{equation}\label{iL:l_as2=}
\tilde{l}(\tilde{t},\tilde{x},\beta_0) \approx
\frac{2(\tilde{t}-\tilde{x})} {\tilde{t}^2-\beta_0^2\tilde{x}^2}
\end{equation}
for $\tilde{t} \gg \gamma_0$ and all $0\leq \tilde{x} \leq \tilde{t}$.
Substituting Eq.~(\ref{iL:l_as2=}) into Eq.~(\ref{iL:dPi_p=}), we
obtain exactly the same equations with respect to
$\tilde{x}_p(\tilde{t})$, $\Pi_p(\tilde{t})$ as those with respect to
$\bar{x}_p(\bar{t})$, $\Pi_p(\bar{t})$ in section~\ref{s:iL0}, which
are then reduced to the same phase equation (\ref{iL0:deta/dhi=}) with
the same topology of integral curves in the vicinity of the singular
point $(\chi,\eta) =(\eta_0,\eta_0)$. As a consequence, we arrive at
conclusions that are fully analogous to those made for the case of
$\Lambda=0$:
\begin{itemize}
\item[(i)] test ions with $\mu> \mu_{\ast} =\frac{1}{8}$ always catch up
and overtake the electron front, being finally accelerated to
$\Pi_{p\infty} >\Pi_0$;
\item[(ii)] for any $\mu< \frac{1}{8}$ there exists a critical value
$\eta_{cr\Lambda}(\mu)$ of parameter $\eta_0$ [or, equivalently, a
critical value $\gamma_{cr\Lambda}(\mu)$ of parameter $\gamma_0$] such
that only for $\eta_0> \eta_{cr\Lambda}(\mu)$ [i.e.\ for $\gamma_0>
\gamma_{cr\Lambda}(\mu)$] can a test ion with the charge-over-mass
ratio $\mu$ overtake the electron front and be accelerated to
$\Pi_{p\infty} >\Pi_0$; for $\eta_0< \eta_{cr\Lambda}(\mu)$ a test ion
lags behind the electron front while its dimensionless momentum
$\Pi_p(\tilde{t})$ asymptotically approaches $\Pi_0$ on a timescale
$\tilde{t}_{ac} \simeq \exp\left(\frac{1}{2}\mu^{-1}\right)$.
\end{itemize}
These conclusions are fully confirmed by numerical integration of
Eqs.~(\ref{iL:dx_p,Pi_p=}) with
$\tilde{l}(\tilde{t},\tilde{x},\beta_0)$ calculated from
Eqs.~(\ref{eL:tx_e=})-(\ref{eL:tt_e1=}).

A remarkable fact is that we have two universal functions, namely,
$\eta_{cr0}(\mu)$ for $\Lambda=0$, and $\eta_{cr\Lambda}(\mu)$ for
$\Lambda> h_0$, which cover the entire range of $\Lambda$ variation.
Since at intermediate times $\tilde{t} \simeq 1$ the two cases of
$\Lambda=0$ and $\Lambda> h_0$ are mathematically not equivalent, the
functions $\eta_{cr0}(\mu)$ and $\eta_{cr\Lambda}(\mu)$ numerically
differ from one another, except for the initial point
$\eta_{cr0}(\mu_{\ast}) = \eta_{cr\Lambda}(\mu_{\ast}) =0$. This
difference, however, is practically not significant, as one verifies by
comparing the numerically calculated values of  $\eta_{cr0}(\mu)$ and
$\eta_{cr\Lambda}(\mu)$ in Table~\ref{t:1}. In the limit of $\mu\ll 1$
one derives an asymptotic expression
\begin{equation}\label{iL:gam_cr=}
  \gamma_{cr\Lambda} = \frac{1}{4} \mu\, \exp\left(
  \frac{1}{\mu}-1\right),
\end{equation}
which is exactly one half of the corresponding limit
(\ref{iL0:eta_cr=}) for $\gamma_{cr0}$. One can safely conjecture that
for intermediate values $0< \Lambda < h_0$ the corresponding critical
values of $\gamma_0$ lie between $\gamma_{cr0}(\mu)$ and
$\gamma_{cr\Lambda}(\mu)$. For $\gamma_0> \gamma_{cr\Lambda}(\mu)$,
when a test ion does overtake the electron front, its final energy in
the limit of $\mu\ll 1$ is given by
\begin{equation}\label{iL:gam_inf=}
  \gamma_{p\infty} = \mu\gamma_0
  \left(1 +\ln\frac{4\gamma_0}{\mu} \right).
\end{equation}

\begin{figure}[htb]
\includegraphics[width=75mm]{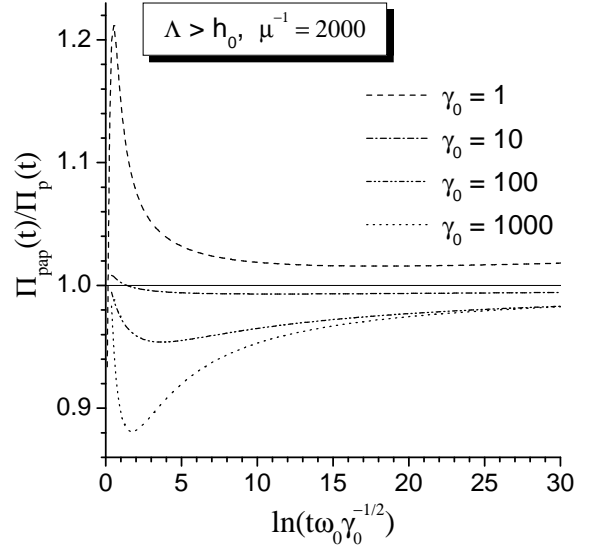}
\caption{Comparison of the intermediate asymptotics
(\ref{iL0:Ias-Pi_p=}) with the results of numerical integration of the
ion equations of motion (\ref{iL:dx_p,Pi_p=}) for $\mu= 1/2000$ and
four different values of $\gamma_0$: plotted is the ratio of the ion
momentum $\Pi_{pap}(t)$, obtained from Eqs.~(\ref{iL0:Ias-Pi_p=}) and
(\ref{iL:Pi_p1=}), to  the numerically calculated value $\Pi_p(t)$
versus time $t$ in units (\ref{eL:[t]=}).} \label{f:9}
\end{figure}

Under the approximation (\ref{iL:l_as2=}) one derives the same
intermediate asymptotics (\ref{iL0:Ias-Pi_p=}) for the test ion
momentum $\Pi_p(t)$  as in the case of $\Lambda=0$. The only difference
between the two cases is in the initial values of $t_{p1}$ and
$\Pi_{p1} = \Pi_p(t_{p1})$ to be used in Eq.~(\ref{iL0:Ias-Pi_p=}).
Unlike in the $\Lambda=0$ case, no appropriate analytical solution to
Eqs.~(\ref{iL:dx_p,Pi_p=}) was found for $\tilde{t} \lesssim 1$ that
would yield suitable expressions for $\tilde{t}_{p1}$ and $\Pi_{p1}$.
It is only in the ultra-relativistic limit $\mu\Pi_0 \gg 1$ that one
obtains a simple result $\Pi_{p1} = \mu\Pi_0$, $\tilde{t}_{p1}
=\left(\frac{1}{2} \mu\gamma_0 \right)^{1/2}$. In the opposite limit of
$\mu\Pi_0 \ll 1$ expansion (\ref{iL:Pi_p(0)=}) suggests that the value
$\Pi_p =\mu\Pi_0$ is achieved at $\tilde{t} \simeq 1$. Taking guidance
from such considerations, we propose simple approximate expressions
\begin{equation}\label{iL:Pi_p1=}
  \Pi_{p1} = \mu\Pi_0, \quad \tilde{t}_{p1} =\left(2+\frac{1}{2} \mu
  \gamma_0 \beta_0^2\right)^{1/2},
\end{equation}
which, on the one hand, agree with the ultra-relativistic limit and, on
the other hand, fit reasonably well the results of numerical
integration of Eqs.~(\ref{iL:dx_p,Pi_p=}) shown in Fig.~\ref{f:9}. As a
result, Eq.~(\ref{iL0:Ias-Pi_p=}) with $\Pi_{p1}$ and $\tilde{t}_{p1}$
taken from Eq.~(\ref{iL:Pi_p1=}) is formally applicable at
$\tilde{t}_{p1} \ll \tilde{t} \ll \exp\left(\frac{1}{2}
\mu^{-1}\right)$. A comparison with numerical results in Fig.~\ref{f:9}
shows that at $\tilde{t} \gtrsim 100$  the intermediate asymptotics
(\ref{iL0:Ias-Pi_p=}) has a typical error of a few percent.

If we consider now the case of a delayed ion start at $ t=t_{p0} >0$,
we find that for $\tilde{t}_{p0} \gg 1$ we have $\tilde{l} \ll 1$ for
all $\tilde{t} \geq \tilde{t}_{p0}$, which again leads us to the
approximation (\ref{iL:l_as2=}) and to the phase equation
(\ref{iL0:deta/dhi=}) with the initial condition $(\chi,\eta) =(0,0)$.
Hence, exactly as in the $\Lambda =0$ case, the critical value of the
$\eta_0$ parameter for $\tilde{t}_{p0} \gg 1$  should be given by the
function $\eta_{cr00}(\mu)$. This conclusion is also fully confirmed by
numerical integration of Eqs.~(\ref{iL:dx_p,Pi_p=}). For $\mu \ll 1$
the dependence of the critical $\eta_0$ on the ion start delay
$\tilde{t}_{p0}$ is as follows: as $\tilde{t}_{p0}$ increases from
$\tilde{t}_{p0}=0$ to $\tilde{t}_{p0}= \sqrt{2}$, the critical value of
$\eta_0$ decreases from $\eta_{cr\Lambda}(\mu)$ to  $\eta_{cr00}(\mu)$,
and for $\tilde{t}_{p0}> \sqrt{2}$ it remains equal to
$\eta_{cr00}(\mu)$.

\subsection{Domain of applicability of the laminar-zone solution}
\label{s:ida}

The foregoing analysis of test ion acceleration has been based on the
assumption that the ion trajectories lie entirely inside the laminar
zone of the electron sheath. As already mentioned in
section~\ref{s:em}, this is true only within a certain domain of our
three-dimensional parameter space $(\mu,\gamma_0, \Lambda)$. Because of
a weak dependence on $\Lambda$, the limits of this domain can be
conveniently analyzed in the two-dimensional $(\mu^{-1},\eta_0)$ plane.

Let $\eta_{lam}(\mu)$ be the threshold value of $\eta_0
=\textrm{arccosh}\, \gamma_0$ at which the trajectory of an ion with a
given charge-over-mass ratio $\mu$ just touches the outer boundary of
the relaxation zone (see Figs.~\ref{f:5} and \ref{f:6}), i.e.\ for
$\eta_0> \eta_{lam}(\mu)$ the ion trajectory lies entirely in the
laminar zone, and for $\eta_0< \eta_{lam}(\mu)$ it penetrates (at least
partially) into the relaxation zone. Typically this touching occurs
near the first local maximum of the corresponding $\chi_r(t)$ curve at
$\bar{t}, \tilde{t} \simeq 15$--20 and lies within the reach of the
TIAC code. Because this early part of the boundary between the two
zones ceases to depend on $\Lambda$ for $\Lambda \gtrsim 2.5$, the same
applies to the function $\eta_{lam}(\mu)$.

\begin{figure}[htb]
\includegraphics[width=75mm]{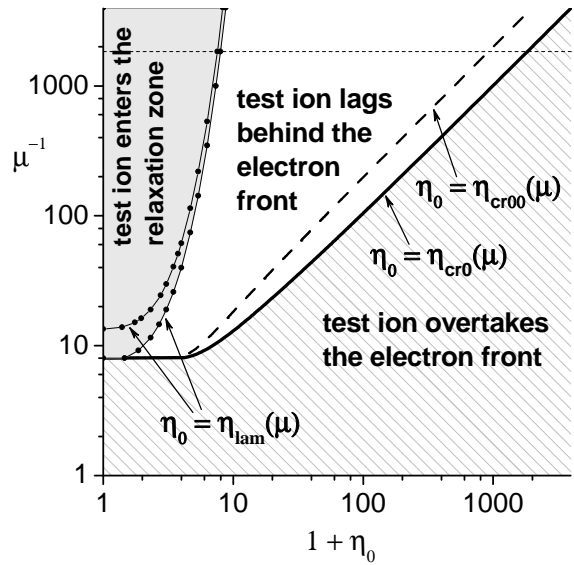}
\caption{Characteristic regions in the $(\eta_0,\mu^{-1})$ parameter
plane. Domain of applicability of the laminar-zone solution lies
outside the grey shaded area delimited by the two $\eta_{lam}(\mu)$
curves, the inner one calculated for $\Lambda=0$, and the outer one for
$\Lambda=10$. Hatched region is where test ions overtake the electron
front: it is delimited by the $\eta_{cr0}(\mu)$ curve which is
practically indistinguishable from $\eta_{cr\Lambda}(\mu)$. Dashed
horizontal line corresponds to the proton value $\mu^{-1} =1836$.}
\label{f:10}
\end{figure}

Figure~\ref{f:10} shows two curves $\eta_{lam}(\mu)$, calculated for
$\Lambda =0$ and 10, which actually span the entire dependence of
$\eta_{lam}$ on the $\Lambda$ parameter. The domain of applicability of
the laminar-zone results lies outside the grey shaded area bounded by
the $\eta_{lam}(\mu)$ curves.  For protons with $\mu= 1/1836$ it
corresponds to $\gamma_0> 348$ at $\Lambda=0$, and to $\gamma_0> 537$
at $\Lambda \gtrsim 2.5$. The fact that the two critical values
$\eta_{cr0}(\mu)$ and $\eta_{cr00}(\mu)$ turn out to be deeply inside
this domain (at least for $\mu< 0.1$) justifies all the conclusions
made in sections~\ref{s:iL0} and \ref{s:iL} about the possibility for a
test ion to catch up with the electron front. Note that the two curves
$\eta_{cr0}(\mu)$ and $\eta_{cr\Lambda}(\mu)$, which span the
dependence of the critical value $\eta_{cr}$ on the $\Lambda$
parameter, are virtually indistinguishable in Fig.~\ref{f:10}.

For protons and heavier ions with $\mu \leq 1/1836$ the following
conclusions can be drawn from Fig.~\ref{f:10}. The position of the
$\eta_{lam}(\mu)$ curves indicates that ion acceleration in the
non-relativistic case of $\Pi_0 \ll 1$  takes place deeply in the
relaxation zone, where one can expect the usual Boltzmann relation to
be a good approximation. The laminar-zone solution is not applicable in
such a case. However, it becomes fully applicable when the electrons
are boosted to highly relativistic energies of $\gamma_0 \gtrsim
300$--500 ($\eta_0 \gtrsim 6$--7).

\section{Conclusion}

In this paper rigorous results are presented for a particular case of
the TNSA mechanism of ion acceleration in a planar electron sheath
evolving from an initially mono-energetic cloud of hot electrons.
Self-consistent treatment of the collisionless electron dynamics fully
captures the effects of departure from the Maxwell-Boltzmann
distribution. These effects come to a foreground in the outer laminar
zone of the expanding electron cloud, where the ion acceleration can be
analyzed by analytical means. In particular, the limiting (in the limit
of $t\to\infty$) gamma-factor $\gamma_{p\infty}$ of an accelerated test
ion can be calculated exactly. Note that the assumption of an
isothermal Boltzmann distribution for hot electrons leads to an
infinite value of $\gamma_{p\infty}$.

It is shown that the limiting value $\gamma_{p\infty}$ is determined
primarily by the values of the two (out of the total three) principal
dimensionless parameters of the problem, the ion-electron
charge-over-mass ratio $\mu =m_eZ_p/m_p$, and the initial gamma-factor
$\gamma_0$ of the accelerated electrons. For $\mu> \mu_{\ast}=
\frac{1}{8}$ a test positive particle (for example a positron) always
overtakes the electron front and reaches $\gamma_{p\infty}> \gamma_0$.
In the physically more interesting case of $\mu< \mu_{\ast}$ the
limiting ion energy depends on whether $\gamma_0$ is above or below a
certain critical value $\gamma_{cr} =\gamma_{cr}(\mu)$, namely, we have
$\gamma_{p\infty}= \gamma_0$ for $\gamma_0< \gamma_{cr}$, and
$\gamma_{p\infty}> \gamma_0$ [as given by Eqs.~(\ref{iL0:ur-gam_inf=})
and (\ref{iL:gam_inf=})] for $\gamma_0> \gamma_{cr}$. Practically
insignificant dependence of $\gamma_{cr}$ on the dimensionless foil
thickness $\Lambda$ is limited to a variation within a factor 2 and
spanned by the functions $\gamma_{cr0}(\mu)$ and
$\gamma_{cr\Lambda}(\mu)$ calculated in Table~\ref{t:1} and
Eqs.~(\ref{iL0:eta_cr=}) and (\ref{iL:gam_cr=}).

For protons and heavier ions with $\mu \leq 1/1836$ we always have
$\gamma_0< \gamma_{cr}$ because the corresponding values of
$\gamma_{cr} \sim \exp(\mu^{-1})$ are enormous and beyond practical
reach. Therefore, in reality these ions can never catch up with the
electron front. Although formally the ion gamma-factor $\gamma_p(t)$ in
this case still tends to $\gamma_0$ as $t \to \infty$, this fact is
also practically irrelevant because $\gamma_p(t)$ approaches $\gamma_0$
on an enormous timescale $t_{ac} \simeq t_{\Sigma}
\exp\left(\frac{1}{2}\mu^{-1}\right)$ [or $t_{ac} \simeq
\omega_0^{-1}\sqrt{\gamma_0}\, \exp\left(\frac{1}{2}\mu^{-1}\right)$
for $\Lambda> 1$] that never occurs in nature. For practical
applications one should use the intermediate asymptotic formula
(\ref{iL0:Ias-Pi_p=}) derived for dimensionless times $1 \ll
\bar{t},\tilde{t} \ll \exp\left(\frac{1}{2} \mu^{-1}\right)$.

Our results for ion motion have been obtained under the condition that
the ion trajectory lies entirely in the laminar zone of the electron
sheath. Numerical investigation of the evolution of the laminar zone
boundaries reveals that this condition imposes a lower bound $\gamma_0>
\gamma_{lam}(\mu)$ on the initial gamma-factor of hot electrons (see
Fig.~\ref{f:10}). The latter inequality reflects a more general fact
that the role of the non-Boltzmann effects in ion acceleration
increases with $\gamma_0$, i.e.\ with the energy of hot electrons. In
particular, acceleration of protons by a non-relativistic electron
sheath occurs practically entirely in the quasi-Boltzmann core of the
electron cloud, where the details of the initial electron energy
distribution are ``forgotten''. However, in the ultra-relativistic case
of mono-energetic electrons with $\gamma_0 \gtrsim 300$--500,
acceleration of protons takes place entirely in the non-Boltzmann
laminar zone of the electron sheath, where full memory of the initial
electron energy distribution has been preserved.

\begin{acknowledgments}
The author gratefully acknowledges stimulating discussions with
M.~Murakami and S.V.~Bulanov.
\end{acknowledgments}


\begin{thebibliography}{99}
\bibitem{ClKr.00} E.L.~Clark \textit{et al.}, Phys.\ Rev.\ Lett.\
\textbf{84}, 670 (2000); \textbf{85}, 1654 (2000).
\bibitem{MaGu.00} A.~Maksimchuk \textit{et al.}, Phys.\ Rev.\ Lett.\
\textbf{84}, 4108 (2000).
\bibitem{SnKey.00} R.A.~Snavely \textit{et al.}, Phys.\ Rev.\ Lett.\
\textbf{85}, 2945 (2000).
\bibitem{HeKa.02} M.~Hegelich \textit{et al.}, Phys.\ Rev.\ Lett.\
\textbf{89}, 085002 (2002).
\bibitem{CoFu.04} T.E.~Cowan \textit{et al.}, Phys.\ Rev.\ Lett.\
\textbf{92}, 204801 (2004).
\bibitem{MoTa.06} G.A.~Mourou, T.~Tajima, and S.V.~Bulanov, Rev.\ Mod.\
Phys.\ \textbf{78}, 309 (2006).
\bibitem{HaBr.00} S.P.~Hatchett \textit{et al.}, Phys.\ Plasmas
\textbf{7}, 2076 (2000).
\bibitem{WiLa.01} S.C.~Wilks \textit{et al.}, Phys.\ Plasmas
\textbf{8}, 542 (2001).
\bibitem{GuPa.65} A.V.~Gurevich, L.V.~Pariiskaya, and L.P.~Pitaevskii,
Zh.\ Eksp.\ Teor.\ Fiz.\ \textbf{49}, 647 (1965) [Sov.\ Phys.\ JETP
\textbf{22}, 449 (1966)].
\bibitem{Mora03} P.~Mora, Phys.\ Rev.\ Lett.\ \textbf{90}, 185002 (2003).
\bibitem{GiJo.86} S.~Gitomer \textit{et al.}, Phys.\ Fluids \textbf{29},
2679 (1986).
\bibitem{CrAu.75} J.E.~Crow, P.L.~Auer, and J.E.~Allen, J.\ Plasma
Phys.\ \textbf{14}, 65 (1975).
\bibitem{PaTi.04} M.~Passoni, V.T. ~Tikhonchuk, M.~Lontano, and
V.Yu.~Bychenkov, Phys.\ Rev.\ E \textbf{69}, 026411 (2004).
\bibitem{PeMo78} J.S.~Pearlman and R.L.~Morse, Phys.\ Rev.\ \textbf{40},
1652 (1978).
\bibitem{KiMi.83} Y.~Kishimoto \textit{et al.}, Phys.\ Fluids \textbf{26},
2308 (1983).
\bibitem{PaLo04} M.~Passoni and M.~Lontano, Laser and Part.\ Beams
\textbf{22}, 163 (2004).
\bibitem{BuEs.04} S.V.~Bulanov \textit{et al.}, Plasma Phys.\ Rep.\
\textbf{30}, 18 (2004).
\end{thebibliography}
\end{document}